\documentclass[]{elsart}
\usepackage{epsfig}
\begin{document}
\begin{frontmatter}
\title{Single-neutron transfer from $^{11}$Be$_{gs}$ via \\ 
       the (p,\,d) reaction with a radioactive beam}

\author[Orsay,Surrey]{J.S. Winfield\thanksref{Catania}}, 
\author[Orsay]{S. Fortier}, 
\author[Surrey]{W.N. Catford}, 
\author[Orsay]{S. Pita}, 
\author[LPC]{N.A. Orr},
\author[Orsay]{J. Van de Wiele},
\author[Orsay]{Y. Blumenfeld},
\author[Paisley]{R. Chapman}, 
\author[Oxford]{S.P.G. Chappell}, 
\author[Birm]{N.M. Clarke},
\author[Surrey]{N. Curtis\thanksref{Florida}}, 
\author[Birm]{M. Freer}, 
\author[Orsay]{S. Gal\`es}, 
\author[Orsay]{H. Langevin-Joliot},
\author[Orsay]{H. Laurent},
\author[Orsay]{I. Lhenry}, 
\author[Orsay]{J.M. Maison},   
\author[Ganil]{P. Roussel-Chomaz}, 
\author[Surrey]{M. Shawcross},
\author[Paisley]{K. Spohr},
\author[Orsay]{T. Suomij\"arvi}
and 
\author[Ganil]{A. de~Vismes}

\address[Orsay]{Institut de Physique Nucl\'eaire, IN2P3-CNRS, 
91406 Orsay cedex, France}
\address[Surrey]{Department of Physics, University of Surrey, Guildford,
Surrey GU2 7XH, UK}
\address[LPC]{Laboratoire de Physique Corpusculaire, IN2P3-CNRS, ISMRA
et Universit\'e de Caen, 14050 Caen cedex, France}
\address[Paisley]{Department of Electronic Engineering and Physics, University 
of Paisley,\\ Paisley PA1 2BE, UK}
\address[Oxford]{Nuclear and Astrophysics Laboratory, University of Oxford,\\ 
Oxford OX1 3RH, UK}
\address[Birm]{School of Physics and Astronomy, University of 
Birmingham, Edgbaston, Birmingham B15 2TT, UK}
\address[Ganil]{GANIL (CEA/DSM-CNRS/IN2P3), BP 5027, 14076 Caen cedex 5, 
France} 

\thanks[Catania]{Present address: INFN - Laboratori Nazionali del Sud, 
  via S. Sofia 44, I~-~95123 Catania, Italy.  E-mail: winfield@lns.infn.it}
\thanks[Florida]{Present address: Physics Department, Florida State 
  University, Tallahassee, Florida 32306, USA.}

\begin{abstract}
  The $^{11}$Be(p,\,d)$^{10}$Be reaction has been performed in inverse
  kinematics with a radioactive $^{11}$Be beam of $E/A = 35.3$~MeV.
  Angular distributions for the $0^+$ ground state, the $2^+$,
  3.37~MeV state and the multiplet of states around 6 MeV in $^{10}$Be
  were measured at angles up to $16^\circ_{cm}$ by detecting the
  $^{10}$Be in a dispersion-matched spectrometer and the coincident
  deuterons in a silicon array.  Distorted wave and coupled-channels
  calculations have been performed to investigate the amount of $2^+$
  core excitation in $^{11}$Be$_{gs}$.  The use of ``realistic''
  $^{11}$Be wave functions is emphasised and bound state form factors
  have been obtained by solving the particle-vibration coupling
  equations.  This calculation gives a dominant 2s component in the
  $^{11}$Be$_{gs}$ wave function with a 16\% [2$^+\otimes$1d] core
  excitation admixture.  Cross sections calculated with these form
  factors are in good agreement with the present data.  The Separation
  Energy prescription for the bound state wave function also gives
  satisfactory fits to the data, but leads to a significantly larger
  [2$^+\otimes$1d] component in $^{11}$Be$_{gs}$.


\end{abstract}


\end{frontmatter}

\newpage

\section{Introduction}

The nucleus $^{11}$Be is of especial interest for several reasons.  As
is well-known, the ground state spin-parity is $1/2^+$ in
contradiction to the simple shell model and spherical Hartree-Fock
prediction of $1/2^-$.  This ``parity inversion'' is correctly
predicted by, for example, recent psd-shell calculations of
Brown~\cite{BAB96}.  The $2s_{1/2}$ intruder orbital is lowered by the
non-central part of the particle-hole interaction~\cite{Tal60}.
Moreover, $^{11}$Be is often regarded as the classic one-neutron halo
nucleus: the small single-neutron separation energy of 505~keV
together with an assumed $s$-wave nature of the valence neutron leads
to a very extended spatial distribution~\cite{Mil83,Han96}.

Several calculations of the $^{11}$Be ground state 
structure have
been performed.  The theoretical approaches include: the shell
model~\cite{War92,Bro98}, the variational shell model~\cite{Ots93},
the Generator Coordinate model~\cite{Des97}, and coupling of the
neutron with a vibrational~\cite{Vin95,Bha97} or rotational
core~\cite{Nun96,Esb95}.  Most of these models correctly reproduce the
parity inversion and high-energy reaction data, but make very
different predictions about the degree of coupling of an s$_{1/2}$
neutron to the $^{10}$Be $0^+$ ground-state core relative to a
d$_{5/2}$ neutron coupled to a $2^+$ excited core (the first excited
state of $^{10}$Be at 3.368~MeV).

A direct test of the models for the structure of $^{11}$Be$_{gs}$ may
be made by measuring the relative cross sections of one-neutron
pick-up reactions feeding the 0$^+$ and 2$^+$ states of $^{10}$Be.
Transfer cross sections depend on the overlap between the wave
functions of the initial and final states through the radial neutron
form factors $u_{lj}$(r).  Standard distorted wave Born approximation
(DWBA) analyses assume that these form factors are proportional to
single particle wave functions $U_{lj}^{sp}$(r), so that one may
calculate cross sections independently of any prior assumption about
the structure of initial and final states, apart from an overall
normalisation factor.  The latter is the spectroscopic factor, which
is defined as the product of the overlap integral $\int
u_{lj}^2(r)r^2{\rm d}r$ and a factor $(n+1)$ \cite{Aus70}, where 
$n$ in the
present case is the neutron occupation number of the 2s1d shell in
$^{10}$Be.  If one expresses the wave function of the 1/2$^+$
$^{11}$Be ground state as the sum of the single particle and core
excited components
\begin{equation}
|^{11}{\rm Be}_{gs}>= \alpha |^{10}{\rm Be}(0^+) \otimes 2{\rm s}> + 
\beta |^{10}{\rm Be}(2^+) \otimes 1{\rm d}>,
\end{equation}
the spectroscopic factors S(0$^+$) and S(2$^+$) 
for transfer to the ground and first excited state of $^{10}$Be 
should be directly related to $\alpha^2$ and $\beta^2$, respectively,
assuming negligible population of the
2s1d orbitals by $^{10}$Be core neutrons.\footnote{Strictly speaking,
$\alpha$ and $\beta$ should be equal to the fractional
parentage coefficients, the squares of which add up to unity.
The relation between these and spectroscopic factors is given in
the Appendix and Ref.~\cite{Aus70}.}  
Table~\ref{tabtheopred} gives 
spectroscopic factors
deduced from the various models cited above. 
These spectroscopic factors
vary widely.  For example,
the standard Shell Model~\cite{War92,Bro98} predicts S(0$^+$) = 0.74
and  S(2$^+$) = 0.19, while the Variational Shell Model~\cite{Ots93}
gives  S(0$^+$) = 0.55
and  S(2$^+$) = 0.40.  
Also given in the table is the ratio 
$R_{ce}$ = S(2$^+$)/[S(0$^+$) + S(2$^+$)],
which provides a measure of the amount of the $2^+$ admixture
in the wavefunction.  This ratio
shows large variations, although all models agree that the 
$2^+$ admixture is smaller than the 2s
component.
The dominance of the 2s component is also the 
conclusion of some recent experimental
investigations~\cite{Gei99,Aum00}.  However,
the precise amount of core-excitation in the  $^{11}$Be ground 
state wavefunction remains uncertain.

\begin{table}
\caption{Coupling of the $^{11}$Be ground state with the
$0^+$ 0.0~MeV and $2^+$ 3.37~MeV 
states $^{10}$Be, as predicted by different structure calculations.
The spectroscopic factors for the square of the
wave function overlaps are given, together with the $2^+$ admixture
$R_{ce}$.}
\begin{tabular}{lccc}
\hline
                  & S(0$^+ \otimes$ 2s) & S(2$^+ \otimes$ 1d) & 
  $R_{ce}$ \\
\hline
Variational shell-model (Otsuka {\it et al.} \cite{Ots93}) & 0.55 & 0.40 
     & 0.42\\
PVM (Bhattacharya \& Krishan \cite{Bha97})     & 0.70 &  -   &  - \\
CC rotational coupling (Nunes {\it et al.} \cite{Nun96}) & 0.78 & 0.20 
     & 0.20\\
Shell-model (Warburton \& Brown \cite{War92,Bro98})  
                                                     & 0.74 & 0.19 & 0.20\\
Vibrational coupling (Vinh-Mau \cite{Vin95,Vin99})   & 0.80 & 0.20 & 0.20\\ 
CC-DC (Esbensen {\it et al.} \cite{Esb95})           & 0.87 & 0.10 & 0.10\\ 
Generator-Coord.\ (Descouvemont \cite{Des97})  & 0.92 & 0.07 & 0.07\\ 
\hline
\end{tabular} \label{tabtheopred}
\end{table}

The present paper describes an approach to the $^{11}$Be$_{gs}$
structure through the $^1$H($^{11}$Be,\,$^{10}$Be)$^2$H
inverse-kinematics radioactive beam reaction. 
Data were also obtained for the $^1$H($^{15}$N,\,$^{14}$N)$^2$H
reaction at 584~MeV for comparison with the (p,\,d) reaction as
measured in normal kinematics~\cite{Sne69}.  Our ($^{15}$N,\,$^{14}$N)
measurement provided an important cross-check of the analysis methods,
particularly in regard to the calculation of the deuteron/heavy-ion
coincidence efficiency and the ability to eliminate background from
reactions on carbon.

Brief accounts of the experiment have been presented in
Refs.~\cite{Lewes98,ENAM98,For99}.  Preliminary data were reported in
Ref.~\cite{ENAM98} together with the first results of a standard DWBA
analysis, which used single-particle form factors evaluated according
to the usual Separation Energy (SE) prescription for the bound state
wave functions.  Spectroscopic factors deduced from that preliminary
analysis~\cite{ENAM98} indicated a large [2$^+\otimes$1d] admixture in
the $^{11}$Be$_{gs}$ wave function, exceeding most theoretical
predictions.  However, that interpretation was valid only to the extent that
the radial wave function $u_{\ell j}$(r) 
of the transferred neutron in $^{11}$Be, relative to a $^{10}$Be core,
may be approximated by the
product of the SE single-particle form factor and a spectroscopic
amplitude.  This assumption is questionable in view of the large
deformation parameter ($\beta_2$=0.74 \cite{Aut70}) of the final
nucleus $^{10}$Be, which could induce important coupling effects.
These core-coupling effects 
may change the radial shape 
of the neutron wavefunction in a manner dependent on the angular momentum.
This can  
directly affect the inferred spectroscopic factor deduced 
from transfer-reaction cross sections.  The latter mainly depend
on the squared amplitude in the surface region of the nucleus.  
In Ref.~\cite{For99} we presented the results of DWBA calculations which
used bound state form factors evaluated in the framework of the
particle-vibration coupling model.  These more sophisticated
calculations indeed gave a significantly smaller [2$^+\otimes$1d]
admixture in the $^{11}$Be$_{gs}$ wave function compared to the
results with the SE wave functions.  In the present paper,  the
vibrational coupling results are
shown to be consistent with calculations which use radial form factors
from the rotational excitation model of Nunes et
al.~\cite{Nun96}. 
In addition, we present calculations which include coupling 
to inelastic channels,
and discuss the most significant factors not included in the
direct transfer calculations.

The structure of the paper is as follows: After the experimental
details (Section~\ref{secExpDetails}), the features of the spectra
are discussed (Section~\ref{secSpectra}).  Next, we explain the
extraction of the cross sections (Section~\ref{secCrxns}), which is
followed by a section on the analysis of the angular
distributions with the 
distorted wave Born approximation (DWBA), including an estimation
of the effect of coupling to inelastic channels
(Section~\ref{secAnalysis}).  
The results are summarised
and discussed in Section~\ref{secConclude}.

\section{Experimental Details} \label{secExpDetails}

\subsection{Beams and targets} \label{secBeamchar}
 
The secondary $^{11}$Be beam was produced by fragmentation of a
65~MeV/nucleon $^{15}$N beam from the GANIL cyclotrons, which
bombarded a carbon target of thickness 1.03~g/cm$^2$ located between
the two superconducting solenoids of the SISSI device~\cite{SISSI}.
The $^{11}$Be beam was analysed by the ``alpha'' spectrometer, which
was operated as a fragment separator and had a 216~mg/cm$^2$ aluminium
achromatic degrader to reject unwanted ion species.  After the
degrader, a small amount of He and Li isotopes remained in the beam
which was 93\% $^{11}$Be.  The magnetic rigidity of the second stage
of the ``alpha'' spectrometer and beam line after the degrader was
2.377~Tm, which corresponded to an average energy of 388.3~MeV
($E/A~=~35.3$~MeV) for the $^{11}$Be ions.  With these settings, the
intensity of the $^{11}$Be particles was about $3 \times 10^4$~pps for
$1.5 \mu$A of $^{15}$N, and the energy spread was 4.0 MeV.  The full
width at half-maximum (FWHM) spread of the incident beam angle at the
reaction target was measured to be
approximately 1.5$^\circ$ horizontally and
0.4$^\circ$ vertically

The $^{11}$Be beam was brought to a dispersive focus at the target,
with a physical size of approximately 5~cm in width and 1.4~cm in
height.  By comparison, the aperture in the target frame was 10~cm by
3~cm.  The principal target used in the experiment was a polypropylene
(CH$_2$CHCH$_3$)$_n$ foil of thickness $50~\mu$m and density
0.896~g/cm$^3$.  The nominal target thicknesses were cross-checked by
comparing the energy-loss of the $^{15}$N direct beam (without
degrading) with stopping-power and range tables~\cite{Zie80}.  Data
were also collected with empty target frames during the course of the
experiment, from which it is deduced that less than
2\% of the counts in the
$^{10}$Be-gated focal plane spectra originated from the target frame
or other sources, rather than from the polypropylene target.  For the
$^{11}$Be beam, data on a 7.24~mg/cm$^2$-thick carbon target were
taken for background subtraction purposes in the singles spectra, as
discussed in Section~\ref{secSpectra}.

For the $^1$H($^{15}$N,\,$^{14}$N)$^2$H reaction, the $^{15}$N primary
beam was degraded in energy by a 745~mg/cm$^2$ carbon foil in SISSI to
584 MeV (38.9 MeV/A), so as to more nearly simulate the properties of
the $^{11}$Be beam.  The resulting spread in the incident beam angle
at the reaction target was about 0.5$^\circ$ horizontally (FWHM) and
0.3$^\circ$ vertically (FWHM), which was somewhat smaller than that
for the $^{11}$Be beam.  The size of the beam spot for the
degraded-energy $^{15}$N at the secondary reaction target was also
smaller than that for the $^{11}$Be, being approximately 4~mm wide and
4~mm tall at half maximum.

\subsection{Dispersion-matched (``energy-loss'') magnetic 
spectrometer} \label{sec:SPEG}

The energy spread of 4~MeV in the incident $^{11}$Be beam, if not
compensated in some way, would make it impossible to separate
$^{10}$Be ejectiles in different excited states.  The necessary
compensation was achieved by the dispersion-matched magnetic
spectrometer, SPEG \cite{Bia89}, in which the initial `analysis'
beamline has a momentum dispersion = 9.86 cm/\%.  The spectrometer
after the target is tuned to match the dispersed beam spot.  In this
way, all particles with the same reaction Q-value can be arranged to
arrive at the same position along the focal plane.

The SPEG spectrometer was placed at a central angle of $0^\circ$, with
the acceptance slits set to $\pm 2.0^\circ$ vertically and $\pm
2.0^\circ$ horizontally.  The focal plane was instrumented to detect
the $^{10}$Be ejectile nuclei.  Momentum and angle measurements were
provided by two XY drift chambers~\cite{Vil89} of the focal plane
detector. This allowed the reconstruction of the position spectrum at
the focal plane, which was at a slight angle to the 
normal to the mean particle
trajectory. Unambiguous particle identification was achieved through a
combination of the energy-loss signal from an ionisation chamber, the
light output from a stopping plastic scintillator, and the
time-of-flight measured by the scintillator relative to the cyclotron
rf.

The unreacted $^{11}$Be beam was stopped in an active ``finger'' of
plastic scintillator material placed at the high momentum side of the
focal plane.  The scintillator was coupled to a phototube and the
light output and time signal were recorded in the acquisition system
by a separate, downscaled trigger.  This allowed the elimination of
the He and Li contamination in the beam, as well as the electrons from
the $\beta^-$ decay of the stopped $^{11}$Be (half-life 13.8~s), and
thus provided a true count of the number of incident $^{11}$Be
particles.

The calibration of the focal plane in terms of the particle momentum
(and hence excitation energy) was performed with a well-collimated
$^{15}$N beam incident on a thin Au target with the spectrometer
centered at $3^\circ$.  The magnetic field of the spectrometer was
changed in 1.5\% steps, two above and two below the central field
setting, and the centroid of the elastic scattering peak recorded.
With the same beam and target, a mask with a pattern of holes was
placed 65~cm after the target for the angle calibration, which was
performed at each of the five field settings used for the momentum
calibration.  The horizontal and vertical angles at the focal plane,
$\theta_{foc}$ and $\phi_{foc}$, respectively, were measured at each
setting.  The horizontal and vertical positions at the focal plane,
$x_{foc}$ and $y_{foc}$, were also measured.  With these calibration
data, the horizontal angle at the target $\theta_{tgt}$ could be
calculated directly from the measured $\theta_{foc}$.  The optical
properties of the spectrometer mean that $\theta_{foc}$ is independent of
$x_{tgt}$ and the other parameters.  However, the vertical focussing
is such that the vertical angle $\phi $ and the vertical position $y$
are mutually dependent.
This is an important effect for the present experiment, noting that
$\phi_{tgt}$ is equally as important as $\theta_{tgt}$ in determining
the reaction angle. Further, the vertical extent of the secondary beam
spot at the target position, $y_{tgt}$\,, is large for a secondary
beam.  To first order, the optical transport of SPEG in the
vertical plane can be written as:
  
\begin{equation}  \label{eq:transopt}
\begin{array}{r@{~=~}l}    
 y_{foc}  & \left({ y \over y}\right)\, y_{tgt} + 
               \left({y \over \phi}\right)\, \phi_{tgt} \\

 \phi_{foc} & \left({\phi \over y}\right)\, y_{tgt} + 
                \left({\phi \over \phi}\right)\, \phi_{tgt} 
\end{array} 
\end{equation}

with the four coefficients depending on $\theta_{tgt}$ and the
momentum difference $\delta = \Delta p/p_0$ measured relative to the
central momentum, $p_0$\,.  These coefficients have been determined
from the angle-mask calibration runs and the microchannel plate
calibration described in Section~\ref{secMCP}.  The measured $y_{foc}$
and $\phi_{foc}$ values were then inverted to determine $\phi_{tgt}$
and the less-important $y_{tgt}$ in the data analysis.  The
uncertainty in $\phi_{tgt}$ was deduced by re-analysing the
calibration runs with the final coefficients applied and was found to
be $\sigma = 0.20^\circ$ for both the $^{11}$Be and $^{15}$N runs,
despite the smaller emittance of the latter beam.

\subsection{Beam tracking detectors}

The angles of the incident beam particles were measured event-by-event 
using two XY position sensitive drift chambers~\cite{Mac98} located 
before the ``analyser''
magnet of the SPEG spectrometer, separated by 1~m from each other.
This position was chosen in order to eliminate the products of any
scattering or nuclear reactions inside the detector material.
Each chamber consisted of four modules such that both X and Y
were measured twice, but in opposite senses (to eliminate multiple 
coincident hits). The approximate 
overall dimension of each chamber was $7 \times 7 \times 7~$cm$^3$.
They were filled with isobutane gas to a pressure of 20~mbar and 
each chamber had a
total thickness of approximately 0.65~mg/cm$^2$ CH$_2$ equivalent,
including  4~$\mu$m of mylar in the gas-window and field-shaping
foils. The timing signals from each module, relative to the focal 
plane plastic scintillator of the spectrometer, were fed into
a multi-hit TDC.  More than one X or Y signal from a given
chamber could thus be recorded 
in a single event (triggered by the spectrometer focal plane).
Correlated
pairs of X and Y times from each chamber 
were then selected by the data analysis software, 
by requiring that the sum of each
pair should add up to a constant.  
If one measurement of
an X-pair was missing in a given chamber, and if the time 
for the X-signal that was present was unique, the X-position
could be recovered knowing that constant, although
this was only required for a few percent of the total events.
The combined efficiency of the two 
chambers for complete XY determination was approximately 88\% 
for the $^{11}$Be beam and  96\%
for the $^{15}$N beam.

Since several magnetic elements in the beam line lay
between the XY-tracking detectors and the target, 
the incident angles at the target 
needed to be cross-calibrated against the measurements at the
tracking detectors.
A tightly-collimated beam was sent
through to the spectrometer focal plane with no reaction target in place.
The currents in two horizontal and vertical steering magnets
were then systematically changed, 
to steer the beam at different angles through the tracking detectors,
and thence through the spectrometer. The difference signals 
$(X_1 - X_2)$ and $(Y_1 - Y_2)$ from the tracking detectors were then
related to the previously-calibrated
$\theta_{tgt}$ and $\phi_{tgt}$ from the focal plane detector.  

\subsection{Microchannel plate} \label{secMCP}

A position-sensitive microchannel plate 
detector~\cite{Odl96} was 
placed approximately 60~cm in front of the 
target and was used as a diagnostic device, as well as to calculate 
the aberration coefficients in Eqs.~\ref{eq:transopt}. It was 
not used for event-by-event correction of the deuteron angle
because of poor detection efficiency.  
However, it was used to
estimate the size of the beam spot, which in turn was used as
input to the Monte-Carlo simulation program 
to calculate the coincidence efficiency as will be discussed in 
Section~\ref{secDetnEffy}.
The horizontal position was calibrated by stepping a tightly-collimated 
beam across the detector using the analysing magnet field.  The vertical
position was calibrated using a dispersed beam
by illuminating a thick target which had a small hole
in the centre, and stepping the target ladder up 
and down.  A further calibration was performed
with an $\alpha$-source by
placing a mask with a pattern of holes immediately in front
of the detector target foil.

\subsection{Detector array for coincident light particles} \label{sec:CHARISSA}

\begin{figure}
\centerline{\psfig{file=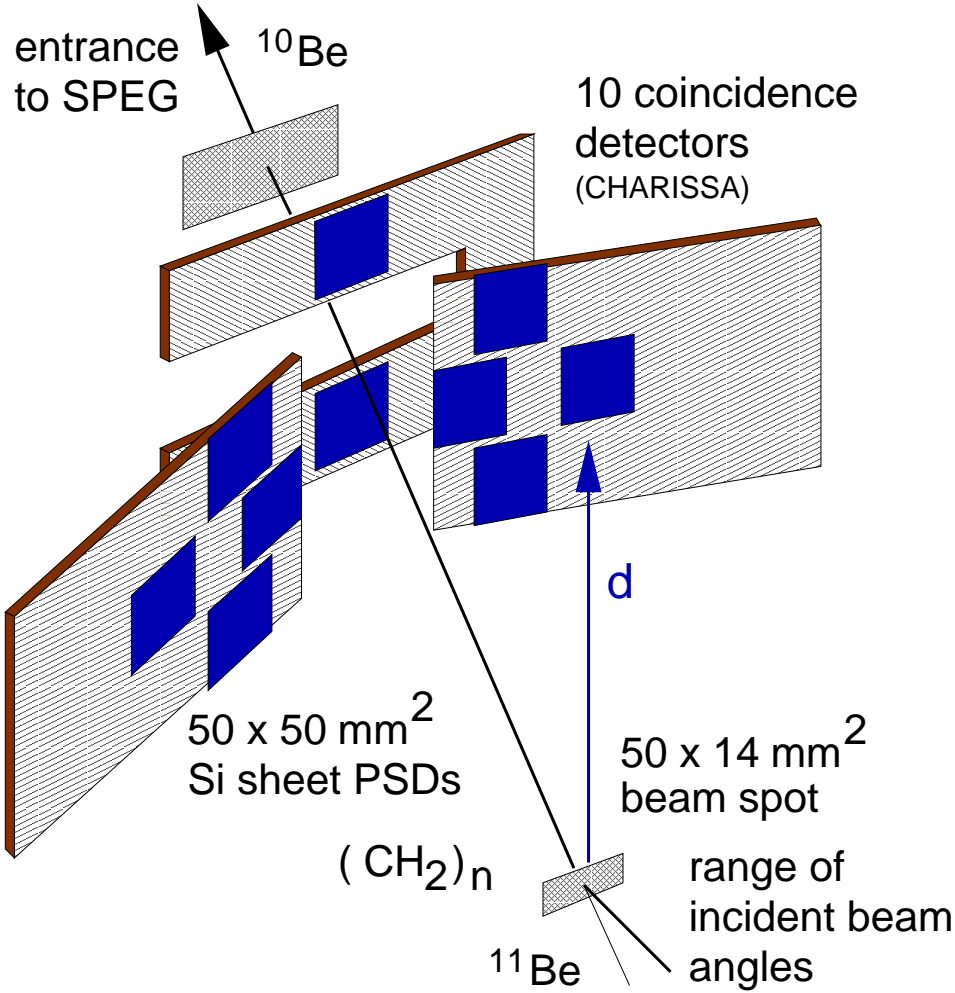,width=0.9\textwidth}}
\caption{Schematic layout of the array of CHARISSA silicon detectors
around the extended target.}
\label{schematic}
\end{figure}

The recoiling 
deuterons were detected in an array of ten
position-sensitive 
sheet-resistive silicon
detectors (CHARISSA) mounted in the  target chamber.  Each detector was 
$5 \times 5$~cm$^2$ in area and 500~$\mu$m thick.  
These are the same detectors as used in the MEGHA array~\cite{Cow99}.
They were arranged around the space joining the extended beam spot and
the spectrometer acceptance aperture, as shown in Fig.~\ref{schematic},
and spanned angles between $5^\circ$ and $35^\circ$ relative to the
central point on the target.  
The energy signals of the detectors were calibrated with a 3-line
$\alpha$-source and the source peaks were monitored during the 
experiment.  The position signals 
were calibrated with the $\alpha$-source, by placing
masks with a pattern of holes over the detector faces.  
The coincidence
timing between the silicon detectors and the spectrometer focal-plane
(taken from the plastic stopping detector) was 
set using the reaction $^1$H($^{15}$N,\,$^{15}$N)$^1$H
at $E$ = 584~MeV
in kinematic coincidence, with the spectrometer moved temporarily
to $-2^\circ$ and each CHARISSA detector moved in turn to  $+75^\circ$.

\begin{figure}
\centerline{\psfig{file=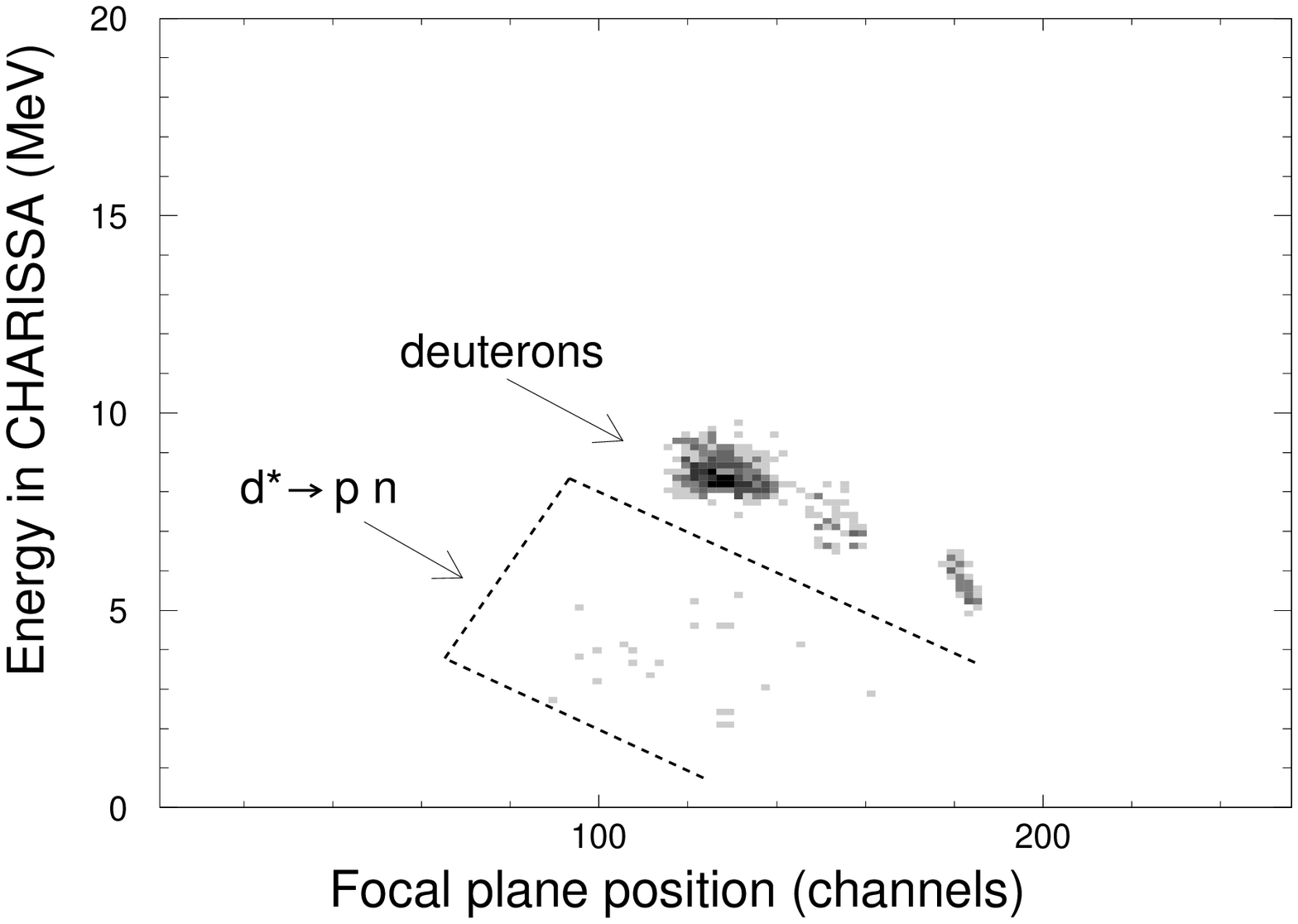,width=0.9\textwidth}}
\caption{Two dimensional plot of  the energy deposited by
particles in the CHARISSA array  against
$^{10}$Be momentum in the focal plane.
The upper groups correspond to deuterons
from the (p,\,d) reaction.  The more diffuse group below the deuterons
corresponds to protons arising from the breakup of excited deuterons.
The data include $^{10}$Be scattering angles from 0.4$^\circ$ to
1.2$^\circ$.}
\label{protonarama}
\end{figure}

The kinematics of the energy signal from the silicon detector 
as a function of the 
momentum of the $^{10}$Be in the focal plane allows
deuterons to be distinguished from protons arising from the
(p,\,d$^* \rightarrow$ p + n) reaction.  This is shown in 
Fig.~\ref{protonarama}.  
It was not possible to use a similar technique to separate the protons
from deuteron breakup in the ($^{15}$N,\,$^{14}$N) data, because the
higher beam energy in that case meant that the deuterons were not in
general stopped in the silicon detectors.

Being less sensitive to the
incident beam angle~\cite{Win97}, the deuteron angle 
proved useful for the beam angle calibration.  Also, the vertical
angle, after the appropriate kinematic transformation to the
heavy-ion angle, could be used as an alternative to $\phi_{tgt}$ 
discussed above. The further use of the deuteron angle was limited
by the large size of the beam spot at the target.

\subsection{Electronics and data acquisition}

Standard NIM and CAMAC electronics were used to process the signals
from the preamplifiers.  
The principal triggers for the
standard GANIL VME-based acquisition system 
were: (i) a SPEG focal plane event, (ii) the downscaled
beam in the plastic ``finger'', and (iii) a CHARISSA ``singles'' event.
The latter was used only as a diagnostic.  The data words for 
event-type (i) included a bit-pattern that indicated which, if any,
CHARISSA detector had recorded a hit in coincidence.
The time correlation was also recorded between the silicon signal
and the heavy-ion in order to reject random coincidences. 
The data were recorded on magnetic tape and were independently 
analysed off-line at the IPN-Orsay and
at the University of Surrey.

\section{Discussion of Focal Plane Spectra} \label{secSpectra}

\subsection{$^{14}$N spectra} 

\begin{figure}
\centerline{\psfig{file=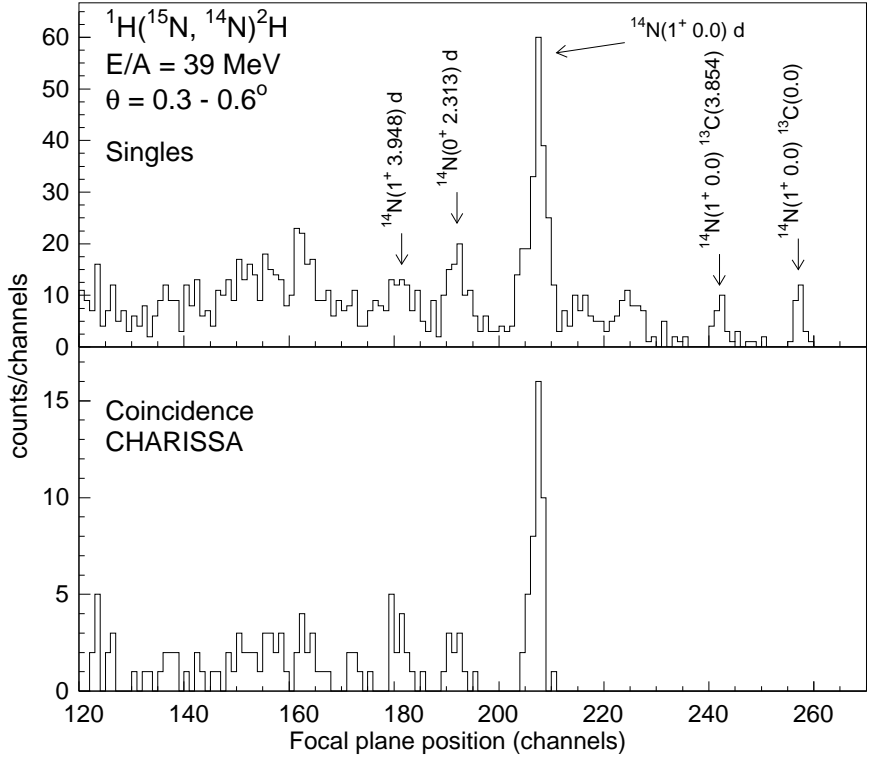,width=0.9\textwidth}}
\caption{$^{14}$N focal plane spectra:  Upper panel in singles, and
  lower panel in coincidence with deuterons in the CHARISSA array.
  The spectra are for $^{14}$N laboratory angles from $0.3^\circ$ to
  $0.6^\circ$.}  \label{14Nspectra}
\end{figure} 
Focal plane position spectra 
for the ($^{15}$N,$^{14}$N) reaction at $E/A = 38.9$~MeV on
the polypropylene target are shown in Fig.~\ref{14Nspectra}.  The
peak from the 
reaction on protons leading to the $^{14}$N $1^+$ g.s.\ dominates the
spectrum.  The experimental energy resolution is about 540 keV (FWHM).
Calculations using the formulae given in Ref.~\cite{Win97} show that
the combined uncertainty in the angle measurement and target effects
contribute about 470 keV to this, leaving a contribution of about 270
keV from the spectrometer optics which corresponds to a momentum
resolving power of approximately one part in 4300.

The coincidence measurement (lower panel) removes the reactions
arising from the carbon in the target, as well as possible background
from $^{15}$N~$\rightarrow~^{14}$N~+~n breakup, seen in the singles
spectrum (upper panel).  Other states observed in $^{14}$N are the
$0^+$ 2.313-MeV level and the second $1^+$ level at 3.948~MeV.  All
these states were seen in the $^{15}$N(p\,,d) experiment of
Ref.~\cite{Sne69}, and with a similar ratio of strength to the present
data.  In the singles spectrum (upper panel), weak yield to low-lying
states in $^{13}$C from the $^{12}$C($^{15}$N,$^{14}$N)$^{13}$C
reaction is observed.  The peak immediately to the left of $^{13}$C
ground state at about 3.9 MeV excitation, and of similar width to the
ground state peak, is assigned to $^{13}$C excited to 3.854~MeV rather
than $^{14}$N excited to 3.948~MeV.  The latter assignment is
unlikely, since the peak would be broadened by inflight $\gamma$-emission
to an estimated
width of about 1.36~MeV.  The excitation energy of $^{13}$C (possibly
mutual with $^{14}$N) for the broad peak near channel 225 is $\sim
9$~MeV.

\subsection{$^{10}$Be spectra} 

The $^{10}$Be focal plane position
spectra for reactions with the $^{11}$Be beam were
accumulated at two different magnetic field settings separated by
0.9\%, for a total number of $3.7 \times 10^9$ incident $^{11}$Be
particles.  Only the data for the first field setting, amounting to
roughly 6/7$^{th}$ of the total, were used in the final analysis.
\begin{figure} 
\centerline{\psfig{file=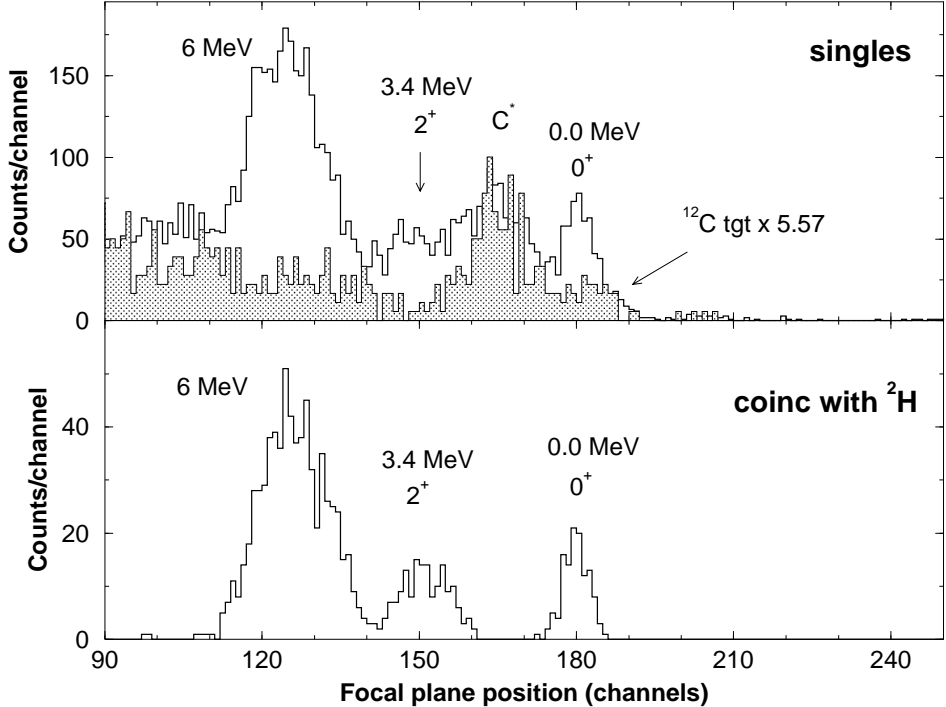,width=0.9\textwidth}}
\caption{$^{10}$Be focal plane spectra. Upper panel: in singles;
  Lower panel: in coincidence with deuterons in the CHARISSA array.
  The spectra are for $^{10}$Be laboratory angles from $0.4^\circ$ to
  $1.2^\circ$.  Superimposed on the upper panel is a spectrum taken on
  a carbon target, normalised to the same number of beam particles and
  equivalent carbon thickness as that for the polypropylene target.
}  \label{BeSpectrum}
\end{figure} 

The ($^{11}$Be,\,$^{10}$Be) singles spectrum is shown in the
upper panel of
Fig.~\ref{BeSpectrum}.
The low-lying states of $^{13}$C from  $^{12}$C in the target
are at most weakly observed; as a reference point, the ground state
for $^{13}$C is expected at channel 222.  In contrast, there is a
strong yield to states at high excitation,
such as the peak at channel 165, corresponding
to an excitation in $^{13}$C of $\sim 10$~MeV, or, more likely, to mutual
excitation of $^{13}$C to $\sim 3.5$~MeV and of $^{10}$Be to 6~MeV.
This strong yield contaminates the peaks from the reaction on
hydrogen.  The carbon-scattering origin of these counts is proved both
by
their absence in the spectrum taken in coincidence with deuterons in
the CHARISSA array (Fig.~\ref{BeSpectrum}, lower panel) and their
presence in
the singles data taken on a pure carbon target
(Fig.~\ref{BeSpectrum}, overlay on upper panel).  The preferential
population of high excitation states in $^{13}$C (with an assumed
$5/2^+$ structure) agrees with Q-value and angular-momentum matching
considerations~\cite{Bri72}.  The coincidence spectrum
(Figure~\ref{BeSpectrum}, lower panel) shows clean separation from
other counts for the ground state and the first excited state
(3.368~MeV, $J^\pi = 2^+$)
of $^{10}$Be. 
As observed in the case of
the $^{15}$N beam spectra, the coincidence measurement very effectively
removes the background which principally arises from the stripping
reaction on the carbon in the target.

The energy resolution for the ground state peak is about 700 keV.  The
angle uncertainty and target straggling accounts for only about 250
keV of this, from which we deduce that the contribution from the
spectrometer optics was some 650 keV.  This corresponds to a momentum
resolving power of approximately one part in 1200, which is
considerably worse than that observed for the ($^{15}$N,\,$^{14}$N)
reaction and is presumably a consequence of the larger beam spot size
in the ($^{11}$Be,\,$^{10}$Be) case.

All the peaks corresponding to
$^{10}$Be excited states in the spectra are broadened
by gamma decay of the nuclei in flight.  The measured
fwhm of the peak for the 3.37~MeV state is $1.14 \pm 0.29$~MeV.  The
maximum spread (i.e.\ base width) from the recoil-broadening is $2
\beta E_\gamma$ = 1.89~MeV, where $\beta = v/c = 0.28$ is the
average $^{10}$Be velocity.  If one assumes an isotropic angular
distribution of $\gamma$-rays, giving a sin($\theta$) dependence to
the spread, the estimated fwhm is 1.26~MeV, in agreement with
observation.  Full treatments of the effect on the broadening when the
angular distribution of the $\gamma$-emission is known are given by
Beene and Devries~\cite{Bee76} and Pelte and Schwalm~\cite{Pel82}.

The dominant peak in the $^{10}$Be spectra arises from transfer to a
group of four closely-spaced levels near 6~MeV excitation.  Kinematic
matching~\cite{Bri72} does not especially favour the 6~MeV region in
$^{10}$Be, thus the strong yield to this region reflects a strong
overlap of initial and final states.  One explanation of the intense
``6-MeV peak'' is the pickup of one p$_{3/2}$ neutron in $^{10}$Be
core, feeding the (s$_{1/2}$,p$_{3/2}^{-1}$) 1$^-$ and 2$^-$ states at
5.960 and 6.263~MeV.  There might also be smaller contributions of 
neutron pickup to the
second 2$^+$ and 0$^+$ states at 5.958 and 6.179~MeV.  However, an
alternative, although not necessarily incompatible explanation invokes
the two-centre shell-model~\cite{Hol71,Fre95}.  In the two-centre
shell-model, the $^{11}$Be nucleus is described as two alpha particles
and three orbiting neutrons~\cite{vonO97}.
The removal of a neutron should leave $^{10}$Be in a similar
two-centre orbit configuration. In the case of $^{10}$Be the most 
strongly-coupled
configurations are not associated with the ground state but with the
states around 6~MeV~\cite{vonO96,Ita00}.  
We note also that the two-centre
shell model can also account for the inversion of the s- and p- states
resulting in the 1/2$^+$ ground state in $^{11}$Be.  This is because of
the lowering of the $\Omega^\pi = 1/2^+$ component of the deformed
d$_{5/2}$ orbitals.

In the singles spectrum, 
the peak at 6~MeV is superimposed on a background
corresponding to the high energy tails of the ($^{11}$Be$\to ^{10}$Be
+ n) and (d$\to$p\,+\,n) break-up reactions near thresholds. The
residual nucleus $^{10}$Be is unbound to particle emission above
6.812~MeV, so no further peaks are expected above this multiplet.  As
previously discussed in Section~\ref{sec:CHARISSA}, the contribution
of the (p,\,pn) reaction to the $^{10}$Be spectra has been removed by
energy conditions in CHARISSA which select coincident recoil
deuterons.

\section{Extraction of Cross Sections} \label{secCrxns}

In this section, we present a description of the simulation program
used to estimate the particle detection efficiencies,
which is needed for the extraction of the 
cross sections from the deuteron coincidence data. 
All the cross sections presented in this
paper are those from the analysis of the coincidence spectra.
Despite the larger number of counts in the peaks, the 
cross section extraction from the singles spectra has no better
statistical accuracy than that from the coincidence spectra, because of
the need to subtract a background which is not accurately determined.
Analyses of some singles spectra have been performed
as a cross-check of the estimation of the coincidence efficiency.

\subsection{Detection efficiency} \label{secDetnEffy}

Data in coincidence with the CHARISSA detectors were corrected for the
variation of geometrical deuteron detection efficiency as a function of
the heavy-ion ($^{10}$Be or $^{14}$N) laboratory angle. This efficiency
was calculated by the use of an extended version of a 
Monte-Carlo simulation program~\cite{Win96},
which accounts for the emittance and beam spot size of the secondary
beam, and for the uncertainty of the scattering angle determination.
Two efficiencies need to be taken into account in the
calculation of the coincidence differential cross section.  Firstly,
the efficiency of detecting deuterons in the CHARISSA array for a
given heavy-ion detection angle.
Secondly, since the beam spot is
large, (particularly for the $^{11}$Be beam), not all the heavy ions
at an angle less than the nominal spectrometer slits setting (which
assumes a point beam spot) will be transmitted.  There is a further
complication arising from the large variation in incident beam angle.
Thus, a second efficiency factor needs to be calculated, which is the
ratio of the number of heavy-ion events generated at a given angle
divided by the number of counts accumulated in the corresponding
``detected'' angle bin.  The
angular resolution of the detection system is also taken into account
in the program.  This latter is an important effect at small angles,
and partly explains the apparently high deuteron
detection efficiency near
$0^\circ$.

\begin{figure} 
\centerline{\psfig{file=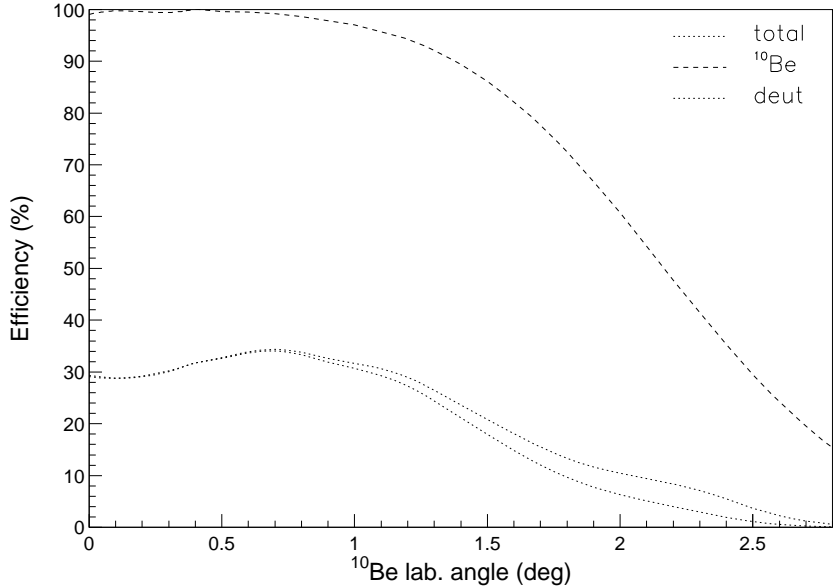,width=0.9\textwidth}}
\caption{Detection efficiency calculated by the Monte-Carlo program
of deuterons in the 
CHARISSA array (dotted line) and of  
$^{10}$Be$_{gs}$
through the spectrometer slits (dashed line)
as a function of $^{10}$Be laboratory angle.  
The solid line is the combined overall coincidence efficiency.
Calculations at discrete angles are plotted, connected by straight lines. }
\label{fig11Be_simul}
\end{figure}

Figure \ref{fig11Be_simul} shows the calculated efficiencies for the
$^{11}$Be beam.  
The beam spot size and envelope 
were set to simulate the experimentally-measured values
given in Section~\ref{secBeamchar}.  

The procedure for the extraction of the 
coincidence cross sections is to integrate the cleanly-separated peaks 
and divide by the total efficiency calculated over the appropriate 
angular range.
For several forward angles in the ($^{11}$Be,$^{10}$Be) data, 
where the background from the carbon target
is relatively small, it was possible to extract the ground
state yield in the singles spectra.  The ratio of the coincident
to singles yield for the $^{10}$Be ions measures the deuteron
deuteron efficiency and 
can be compared to the one calculated
by the simulation program.  For an angular range between 
2.5 to $8.5^\circ_{cm}$,
where the calculated efficiency curve is roughly constant with angle
(Fig.~\ref{fig11Be_simul})
the average coincident to singles yield ratio was found to be
$0.28 \pm 0.04$.  This agrees within errors with the average
efficiency of 31.4\% calculated by the simulation program.

\subsection{$^{14}$N Angular Distributions} \label{sec14Nangdisexp}

\begin{figure}[h]
\centerline{\psfig{file=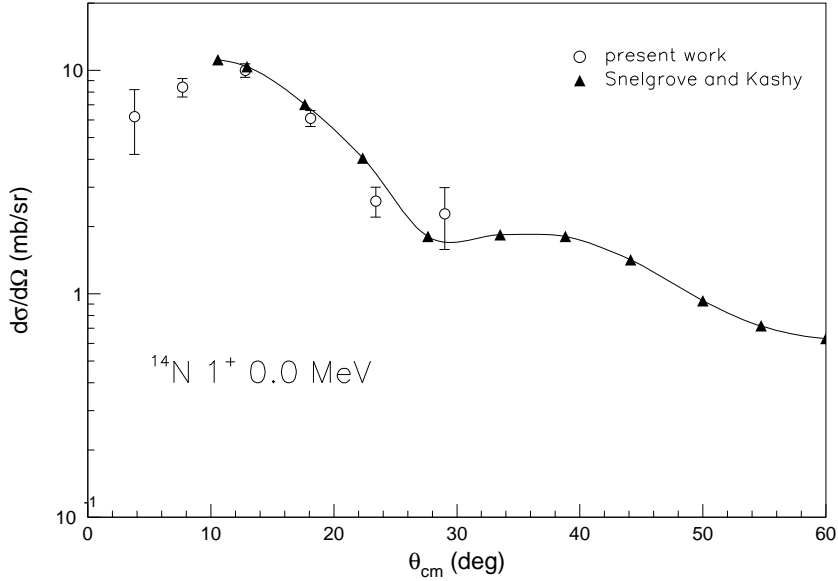,width=0.9\textwidth}}
\caption{Experimental angular distributions for the present inverse
  kinematics reaction to the ground state in $^{14}$N (open circles)
  compared to the (p,\,d) data of Snelgrove and Kashy~{\protect
  \cite{Sne69}} (triangles).  The curve is to guide the eye only.}
\label{angdist_14N_data}
\end{figure}

Cross sectional angular distributions for the 1$^+$ ground state of
$^{14}$N are shown in Fig.~\ref{angdist_14N_data}.  These are from the
analysis of the deuteron-coincident spectra.  Other states in $^{14}$N
were not analysed because of the low yield and the uncertainty in
subtracting the underlying background.

The last angle bin extracted is for $\theta_{lab} = 2.1 - 2.3^\circ$,
corresponding to a mean angle of $\theta_{cm} \approx 29^\circ$.  In
principle, one could go beyond the nominal angular acceptance limit of
the spectrometer ($\pm 2^\circ$ square for a point beam spot, giving a
diagonal limit of $2.83^\circ$) because of the spread in incident
beam angle.  However, the fall-off in reaction cross section and lack of
coincidence efficiency at large angles prevents this.

In Fig.~\ref{angdist_14N_data} we also show the (p,\,d) data of 
Snelgrove and Kashy~\cite{Sne69}, which were measured at $E_p = 39.8$~MeV. 
Our inverse-kinematic data at $E/A = 38.9$~MeV are in 
reasonable agreement with
the (p,\,d) data. 
We conclude 
that our analysis technique in general, and our simulations
of the deuteron efficiency in particular, are reliable 
at least up to a
centre-of-mass angle of $24^\circ$ for p($^{15}$N,\,$^{14}$N)d.  For
that reaction, $24^\circ_{cm}$ corresponds to a deuteron laboratory
angle of approximately $28^\circ$.  If we take the latter as a limit
of reliability on the deuteron scattering angle, and apply it to the
p($^{11}$Be,\,$^{10}$Be$_{gs}$)d reaction, we obtain a rough limit of
$15^\circ_{cm}$ up to which we are confident in the efficiency
simulation.  The 
corresponding limits for the reaction leading to the 3.368~MeV and
6.1~MeV excited states of $^{10}$Be are $18^\circ_{cm}$ and
$21^\circ_{cm}$, respectively.

\subsection{$^{10}$Be Angular Distributions}

The final experimental angular distributions for the 0$^+$ and 2$^+$
states and the 6 MeV multiplet are shown in
Fig.~\ref{angdist_10Be_DWBA} and tabulated in
Table~\ref{tab:xsec_10Be}.  The error bars are statistical.  For the
four most-forward angle points, the systematic errors arising
from uncertainties on the angle determination in SPEG and in the
CHARISSA detectors, and on the variation efficiency of the detection
system as function of the scattering angle, are estimated to be of
the order of 15\%.  
The largest angle points have been
revised from the data shown in ref.~\cite{For99} as a result of a new
estimation of the coincidence efficiency~\cite{Pit00}.  These
largest-angle points are close to
the limits of the reliability criterion
deduced from the ($^{15}$N,\,$^{14}$N) data above, and are more
sensitive to the approximations made for the beam profile, giving an
estimated $25-30$\% 
additional systematic uncertainty from the
efficiency calculation.  The largest-angle points were not used in the
extraction of spectroscopic factors in ref.~\cite{For99}, neither are
they so used in the present paper.
   
\begin{table}
\caption{Differential angular distributions for  $^{11}$Be(p,\,d) leading
to states in  $^{10}$Be.}
\label{tab:xsec_10Be} 
\begin{tabular}{r r c r r c r r}
\hline 
\multicolumn{2}{c}{$0^+$ 0.00~MeV} & & \multicolumn{2}{c}{$2^+$ 3.34~MeV} 
                                   & & \multicolumn{2}{c}{6~MeV states} \\
\cline{1-2} \cline{4-5} \cline{7-8}
$\theta_{cm}$ &  $\sigma(\theta)$ & &
$\theta_{cm}$ &  $\sigma(\theta)$ & &
$\theta_{cm}$ &  $\sigma(\theta)$     \\
 (deg)         &    (mb/sr)        & &
 (deg)         &    (mb/sr)        & &
 (deg)         &    (mb/sr)            \\
\hline 
   2.50         &   $5.21 \pm 1.39$ & &
   2.62         &   $4.85 \pm 1.30$ & &  
   2.73         &  $19.54 \pm 2.54$ \\    
   5.08         &   $5.16 \pm 0.73$ & &
   5.39         &   $6.03 \pm 0.74$ & &
   5.62         &  $20.85 \pm 1.32$ \\
   8.27         &   $4.05 \pm 0.53$ & &
   8.66         &   $4.87 \pm 0.55$ & &
   9.00         &  $17.80 \pm 0.98$ \\
   11.60        &   $1.30 \pm 0.27$ & &
   12.15        &   $4.74 \pm 0.49$ & &
   12.63        &  $22.08 \pm 1.04$ \\
   15.02        &   $0.55 \pm 0.25$ & &
   15.74        &   $4.57 \pm 0.79$ & &
   16.39        &  $19.47 \pm 1.42$ \\
\hline 
\end{tabular}
\end{table}

\section{Analysis of Angular Distributions} \label{secAnalysis}

\subsection{Optical model potentials} \label{secOMP}

Different combinations of optical potentials for the entrance and exit
channels have been tried in the calculations presented below, in order
to test the sensitivity of the extracted spectroscopic factors to the
input parameters.  All the optical potentials used in the present
analysis have the standard Woods-Saxon or Woods-Saxon derivative form.

For the entrance channel, three principal optical potentials have
been used.
The most-recent global nucleon-nucleus optical parameterisation is the
``CH89'' one of Varner \etal~\cite{Var91}.  This has dependences on
energy, mass and isospin, adjusted for a range of stable nuclei from
masses $A = 40$ to 209.  However, data from recent proton elastic 
scattering
experiments involving $^{10}$Be and $^{11}$Be radioactive beams (at
similar incident energies to the present work) 
were only reproduced with the CH89 parameterisation
if the depth of the real Saxon-Woods well were reduced by a factor
of 0.88~\cite{Lap98}.  The CH89-derived
potential $P_1$ in Table~\ref{tab_OMP}
corresponds to p~+~$^{11}$Be at 35.3 MeV with the real well depth
adjusted by this factor.  Secondly, the 1980 parameterisation of
Fabrici et al.~\cite{Fab80} was used to generate the proton potential
$P_2$, assuming a deformation parameter for $^{11}$Be of zero.
Thirdly, 
the proton-nucleus potential
$P_3$ was obtained from the parameterisation of Watson, Singh and
Segel~\cite{Wat69}, which was derived from the analysis of elastic
scattering data in the 1p shell.\footnote{We note that the spin-orbit
parameterisation in Ref.~\cite{Wat69} is non-standard, having
$1/R_{so} ~{\rm d}f / {\rm d}r$ instead of the usual 
$1/r ~{\rm d}f/{\rm d}r$.  
Since the spin-orbit potential has little effect on
the angular distributions, 
we have nevertheless kept to the standard form.}

\begin{table} 
\caption{Optical model parameters for the p + $^{11}$Be entrance 
($P_n$) and d + $^{10}$Be exit ($D_n$)
channels.  Potentials $D_2$ and $D_3$ are 
adiabatic potentials taken from the nucleon potentials of 
Ref.~{\protect \cite{Var91}}
and  {\protect \cite{Wat69}}, respectively,
for use in the Johnson-Soper approximation.
Potential depths are in MeV while diffuseness and reduced radii
are in fm.  Radii are defined in terms of the reduced radii as 
$R_x = r_x \times A^{1/3}$.
}
\begin{tabular}{ccllllllllll} \hline
Label & Ref          & $V$  & $r_v$ & $a_v$ & $W_v$ & $W_s$ & $r_w$ & $a_w$ 
 & $V_{so}$ & $r_{so}$ & $a_{so}$ \\  
\hline
$P_1$ & \cite{Var91} & 41.0 & 1.15  & 0.69  & 3.64  & 7.77  & 1.14  & 0.69 
 & 5.9 & 0.80 & 0.63 \\ 
$P_2$ & \cite{Fab80} & 56.0 & 1.08  & 0.64  & 3.70  & 3.91  & 1.25  & 0.68 
 & 5.6 & 1.01 & 0.60 \\ 
$P_3$ & \cite{Wat69} & 58.4 & 1.12  & 0.57  & $-$   & 10.39 & 1.12  & 0.50 
 & 5.5 & 1.12 & 0.57 \\ 
$D_1$ & \cite{Per63} & 76.8 & 1.15  & 0.81  & $-$   & 23.04 & 1.34  & 0.68
 & $-$ &      &      \\ 
$D_2$ &  a)          & 73.8 &  1.15 & 0.72  & 4.08  & 15.83 & 1.14  & 0.72 
 & 5.9 & 0.78 & 0.63 \\ 
$D_2$ &  b)          & 74.8 &  1.15 & 0.72  & 3.71  & 16.17 & 1.14  & 0.72 
 & 5.8 & 0.78 & 0.63 \\ 
$D_3$ &   a)        & 106.3 & 1.13  & 0.60  & $-$   & 15.9 & 1.13  & 0.53
 & 5.5 & 1.13 & 0.57 \\ 
$D_3$ &   b)        & 107.3 & 1.14  & 0.60  & $-$   & 16.1 & 1.14  & 0.53
 & 5.5 & 1.13 & 0.57 \\ 
\hline
\end{tabular} \label{tab_OMP}
\footnotesize{a) adiabatic potential for d~+~$^{10}$Be$_{gs}$.}\\
{\footnotesize b) adiabatic potential for d~+~$^{10}$Be$_{2^+}$.}\\
\end{table}

The situation for suitable exit channel scattering potentials is not
as satisfactory as that for the entrance channel.  A ``global''
parameterisation may be found in Ref.~\cite{Per63}, but it has been
derived from the analysis of deuteron scattering on a limited number
of $A > 24$ targets.  No polarised beam data were available at the
time that the analysis was made, and so a spin-orbit potential is
lacking.  This potential, adapted for 36~MeV deuterons, is listed as
$D_1$ in Table~\ref{tab_OMP}.  Zwieglinski \etal~\cite{Zwi79} also
used the Perey and Perey~\cite{Per63} parameterisation in the analysis
of their $^{10}$Be(d,\,p)$^{11}$Be data but added their own spin-orbit
potential (depth 7~MeV).  Test calculations that we have performed
with such an added {\it ad-hoc} spin-orbit potential to $D_1$ gave
only small differences to the cross sections in the present, rather
restricted, angular range of interest.

\subsection{Single-particle form factors for a Woods-Saxon well} 
  \label{sec:SEcalcs}

Theoretical differential cross sections $\sigma_{DW}$ have been 
calculated
using the zero-range DWBA code DWUCK4~\cite{DWUCK}.  Test calculations
were made with corrections for the effects of finite range and the
non-locality of optical potentials; in these tests, a finite range
parameter of 0.621 and non-locality ranges of 0.85 and 0.54, for
nucleons and deuterons, respectively, were used.  Differences in the
cross sections with and without the finite range correction were found
to be less than 10\%. 
The effect of the non-locality correction on the
spectroscopic factors was found to be generally less than 20\%.
For some particular optical potentials, the cross sections
changed by about 30\%, 
although in these cases the effect on the
$0^+$ and $2^+$ cross sections is correlated to some extent.  The
DWUCK4 calculations presented here have no non-locality corrections.
Transitions to the 0$^+$ ground state and 2$^+$ first excited state
were assumed to proceed through neutron pick-up in the 2s$_{1/2}$ and
1d$_{5/2}$ orbitals, respectively.  The doublet of 1$^-$ and 2$^-$
states around 6 MeV was assumed to be excited through a p$_{3/2}$
neutron pick-up from the $^{10}$Be core of $^{11}$Be.

For the calculations described in this section, bound state neutron
form factors were calculated in a Woods-Saxon well according to the
usual Separation Energy (SE) prescription~\cite{Aus70}. This consists
of adjusting the well depth so that the eigenvalue of the
Schr\"odinger equation is equal to the experimental separation energy,
thus ensuring correct asymptotic behaviour of the wave function.
In the calculations a spin-orbit Thomas term was used with
$\lambda$=25 and two different geometries for the Woods-Saxon-well:
(1) standard geometrical parameters $r_0$=1.25 fm, $a$=0.65 fm, and
(2) $r_0$=1.15 fm, $a$=0.57 fm, from the parameterisation of
Ref.~\cite{Wat69} for p-shell nuclei.

Different sets of optical potentials were used in the generation of the
distorted waves in the entrance and exit channels, as detailed in
Section~\ref{secOMP}.  Attempts to increase the radius parameter of $P_2$
and $P_3$ by 25\% 
to account for the particularly large matter radius
\cite{tani88} of $^{11}$Be were not successful in reproducing the
present angular distributions, and the geometry given in
Refs.~\cite{Fab80,Wat69} was adopted for the production run
calculations.

Two fundamentally-different types of deuteron potentials have been
used.  Potentials deduced from elastic scattering analyses, such as
the Perey potential $D_1$~\cite{Per63}, are suitable for standard DWBA
calculations.  On the other hand, the adiabatic deuteron breakup
approximation (ADBA) proposed by Johnson and Soper~\cite{Joh70} is
generally known to improve the description of (p,d) and (d,p)
reactions, by accounting for the effects arising from the break-up of
the deuteron in the nuclear field.  Such adiabatic deuteron potentials
were obtained by folding the neutron and proton global potentials at
half the deuteron energy, according to the prescription of
Ref.~\cite{Sat71}. Potential $D_2$ was derived from the
CH89~\cite{Var91} parameterisation and potential $D_3$ from the
Watson, Singh and Segel~\cite{Wat69} parameterisation.

\begin{figure}
\centerline{\psfig{file=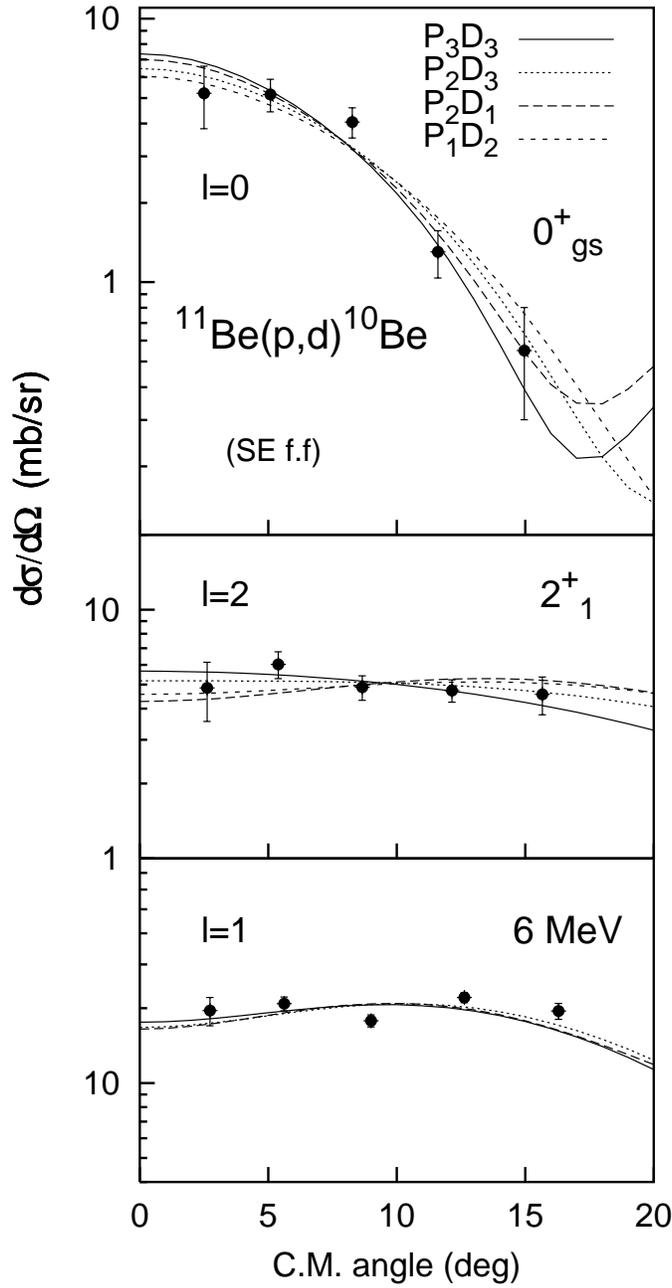,angle=-90,width=0.9\textwidth}} 
\caption{Theoretical angular distributions calculated under the DWBA
  obtained with single-particle SE form factors for states in
  $^{10}$Be.  The points are the experimental angular distributions.}
\label{angdist_10Be_DWBA}
\end{figure}

These different combinations of optical parameter sets reproduce
reasonably well the rapid decrease of the ground state $l$=0 cross
section with angle and the rather flat angular distribution
observed for the 2$^+$ state and the 6 MeV peak (see examples in
Fig.~\ref{angdist_10Be_DWBA}).  The
data in general were poorly reproduced with the global potential
\cite{daehnick80} extrapolated from (d,\,d) scattering on $A \ge 27$
nuclei, or the potentials used for analysing $^{10}$Be(d,\,d) data at
12 and 15 MeV \cite{Aut70}.  (The results from these ``failed''
calculations are neither shown, nor are the potentials considered
further.)  The calculated angular distributions for a given $lj$
transfer were normalised to the four most forward angle data points by
a least square fit procedure, in order to determine the spectroscopic
factors $S$.  These factors were deduced from the relation $\sigma
_{exp}=N S \sigma_{DW}/(2j+1)$, where $j$ is the total spin transfer,
with the (p,\,d) zero range normalisation factor
$N$ of 2.29~\cite{DWUCK}.  The square of the isospin Clebsch-Gordan,
$C^2$, which is sometimes included in the definition of $S$, is unity
for $^{11}$Be$_{gs} \rightarrow ^{10}$Be + n.

\begin{figure}
\centerline{\psfig{file=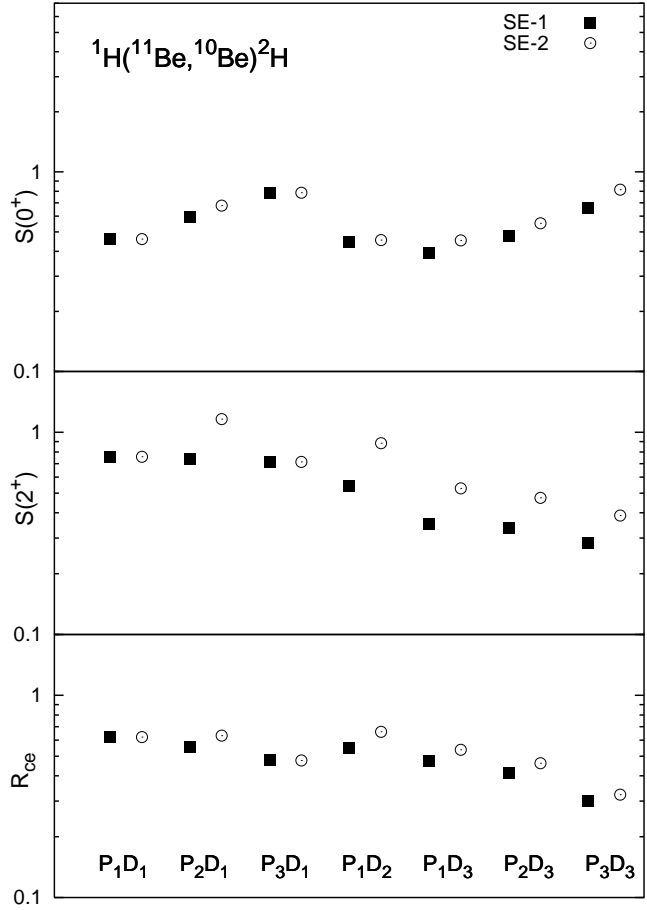,height=14cm}} 
\caption{Spectroscopic factors and the ratio 
  $R_{ce}$ = S(2$^+$)/[S(0$^+$)+S(2$^+$)], extracted from a standard
  DWBA analysis, using different optical parameter sets and
  single-particle form factors SE-1 and SE-2, calculated with
  geometrical parameters (1) and (2), respectively.  }
\label{spectfact_DWBA}
\end{figure}

The spectroscopic factors S($0^+$) and S($2^+$) are shown in
Fig.~\ref{spectfact_DWBA} for various sets of optical potential
parameters.  The experimental S(2$^+$) values obtained by assuming a
pure d$_{5/2}$ neutron pick-up change by less than a few per cent if
instead an $l=2$ neutron pick-up in the higher-lying d$_{3/2}$
subshell is considered. The absolute spectroscopic factors strongly
depend on optical parameters, with for example a factor of about 1.8
between the S(2$^+$) values obtained with the Perey and Perey deuteron
potential $D_1$ and the adiabatic potential $D_3$. The sum of
spectroscopic factors for 0$^+$ and 2$^+$ states, related to the
occupation number of the s-d shell in $^{11}$Be$_{gs}$, was found to
vary between the extreme values of 0.74 and 1.80, with an average
value of 1.09.  On the other hand, the ratio
$R_{ce}$~=~S(2$^+$)/[S(0$^+$)+S(2$^+$)] related to the admixture of
core excited components in the wave function, is less dependent on the
calculation parameters, as shown in Fig.~\ref{spectfact_DWBA}.

The average value of $R_{ce}$ deduced from this SE analysis is
0.51, with a standard deviation of 0.09.  Note that the lowest values
of $R_{ce}$ are obtained from the ADBA calculations with the optical
parameter set $P_3D_3$.  These parameters also provide the best fit to
the data for angles below $~14^\circ_{cm}$, and when SE form factors
are used with either geometry (1) or (2) for the binding potential,
one obtains $R_{ce}$ values of 0.30 and 0.32, respectively.  We
therefore deduce a conservative lower limit of 30\% core excitation
admixture in the $^{11}$Be$_{gs}$ wave function from the analysis
which used single-particle form factors calculated in a Woods-Saxon
well with the Separation Energy procedure.

The angular distribution for the 6~MeV peak is well reproduced by
calculations for $l=1$ transfer to the 1$^-$ and 2$^-$ states at 5.96
and 6.26 MeV (see Fig.~\ref{angdist_10Be_DWBA}), without consideration
of a possible coupling of $^{11}$Be$_{gs}$ to the 2$^+_2$ and 0$^+_2$
states unresolved in the multiplet.  With the optical model parameter
set $P_3D_3$ and form factor geometry (1), the summed spectroscopic
factor is found to be 1.40.  This summed spectroscopic factor is in
good agreement with the results of the shell model calculations of
Warburton and Brown~\cite{War92,Bro98} in an extended 1p-2s-1d basis,
which predict spectroscopic factors of 0.69 and 0.58 for calculated
1$^-$ and 2$^-$ states at 5.96 and 6.23 MeV.  However, we
recall that this parameter set, $P_3D_3$, gave the lowest spectroscopic
factors for the transition to the $^{10}$Be 3.37-MeV
state (see fig.~\ref{spectfact_DWBA}).  Calculations for the 1$^-$
and 2$^-$ states with parameter set $P_3D_1$ gave a summed
spectroscopic factor of 2.69, which is twice the shell model
predictions.  On the other hand, the shape of the experimental 
angular distribution suggests that the amount of any $l = 0$ or $l = 2$
contribution is quite small.  The agreement of the 
$P_3D_3$ calculations with the shell model predictions is thus
circumstantial evidence supporting that choice of optical model 
potentials.

\subsection{Effects of core coupling on transfer cross sections}

The comparison in the previous section between experimental and
theoretical spectroscopic factors is based on two assumptions: (i)~a
single-particle bound state form factor, and (ii)~negligible
interference from two-step processes.  We have noted in the
Introduction the particularly large experimental deformation parameter
$\beta_2$=0.74~\cite{Aut70}, of the 2$^+$ state in $^{10}$Be as deduced
from (p,p$^\prime$) data. This might induce significant coupled
channel effects on the (p,d) cross sections.  In addition, the SE
assumption of the proportionality of the bound state form factors to
the single-particle wave functions may be a poor approximation if
certain basic conditions are not fulfilled.  These conditions
include having an experimental binding energy close to that expected
from the simple shell model and negligible residual interaction
between the core and the transferred nucleon~\cite{Aus70}.
In the case of the pick-up of the halo neutron orbiting
around a $^{10}$Be core, coupling to the excited core 
may also modify the radial shape of the neutron
wave function in a significant way.


A description of the theoretical formalism and of the calculation of
the vibrational form factors is given in the Appendix 
(further details will be given in
Ref.~\cite{Pit00}).  Briefly, the $^{11}$Be$_{gs}$
wave function was written in the form given in eq.~(1), and a set 
of coupled equations were solved for the nucleon wave functions.
Calculations were performed for both sets of Woods-Saxon well
geometries (1) and (2) previously adopted for the SE calculations, 
and for two different conditions on the deformation parameter value
$\beta_2$.  By adjusting the strength of the real central and
spin-orbit potential, the eigenvalues of the coupled equations 
were made
equal to the experimental separation energies.

For the VIB-1 and VIB-2 cases listed in Table~\ref{tab:VIBpar}, the
experimental $\beta_2R$ value of 1.84 fm from the analysis of
$^{10}$Be(p,\,p$^\prime$) data \cite{Aut70} was used.  The values of
the potential depths are given in Table ~\ref{tab:VIBpar}.  Note the
unusually large $V_{so}$ (about twice the standard spin-orbit value)
needed to reproduce the $^{11}$Be$_{gs}$ parity inversion.  A similar
result was found by Nunes et al.~\cite{Nun96}.  In spite of the
different radial shapes of $u_{lj}$(r), similar theoretical
spectroscopic factors S$_{s_{1/2}}^{th} \approx 0.84$ and
S$_{d_{5/2}}^{th} \approx 0.16$ are deduced from both calculations.

\begin{table}
\caption{Parameters and resulting spectroscopic factors 
of vibrational coupling calculations for $^{11}$Be.  
S$_{p_{1/2}}$ measures the overlap 
between $^{10}$Be$_{gs}$ and  1/2$^-$ state at 0.32 MeV.
S$_{s_{1/2}}$ and S$_{d_{5/2}}$ are for
transitions to the 0$^+$ ground state and first 2$^+$
state of $^{10}$Be, respectively. S$^{exp}$ values were obtained 
from the present DWBA analysis with optical parameter set $P_3D_3$,
using the normalised $U_{lj}$ wave functions as form factors.
Numbers given in parentheses correspond to the results obtained
using standard SE form factors in the same well geometry.
Potential depths are in MeV and distances in fm.}
\begin{tabular}{ccccccccccc} \hline
&$\beta_2$& $V_0$& $V_{so}$& $r_0$ & $a$ & S$_{p_{1/2}}^{ th}$&
S$_{s_{1/2}}^{ th}$&
S$_{s_{1/2}}^{ exp}$& S$_{d_{5/2}}^{ th}$& S$_{d_{5/2}}^{ exp}$\\
\hline
 VIB-1  & 0.68& 51.5& 15.4& 1.25& 0.65& 0.84& 0.84& 0.67& 0.16& 0.17\\
 VIB-1b & 0.92& 51.0&  7.0& 1.25& 0.65& 0.61& 0.89& 0.65& 0.09& 0.13\\
{\it SE-1} &  &     &     & 1.25& 0.65&     &     &(0.66)&    &(0.28)\\
 VIB-2  & 0.74& 60.3& 13.3& 1.15& 0.57& 0.81& 0.83& 0.80& 0.16& 0.26\\
 VIB-2b & 0.90& 60.0&  8.3& 1.15& 0.57& 0.67& 0.87& 0.79& 0.11& 0.21\\
{\it SE-2} &  &     &     & 1.15& 0.57&     &     &(0.79)&    &(0.38)\\
\hline
\end{tabular} \label{tab:VIBpar}
\end{table}

A smaller 
core-excitation (S$_{d_{5/2}}^{th} \approx 0.10$), is predicted by the
calculations VIB-1b and VIB-2b, which have a more standard, smaller
spin-orbit strength, corresponding to a Thomas-Fermi $\lambda$ factor
of 25.  This requires an increased deformation parameter $\beta_2$ of
approximately 0.9 to reproduce the experimental binding energies.

\subsubsection{Direct one-step transfer}

Theoretical cross sections for direct transfer using external form
factors were calculated with the code DWUCK4 \cite{DWUCK}.  The use of
vibrational form factors (normalised to unity) in place of the SE ones
in DWBA or ADBA calculations neither appreciably modifies the shape of
angular distributions nor the magnitude of the 0$^+$ cross section,
but enhances the theoretical cross section for the 2$^+$ state by
typically a factor of two.  As shown in Ref.~\cite{For99}, the effect
of the coupling on the radial wave functions is only minor for
s$_{1/2}$, but significantly shifts the d$_{5/2}$ form factor
outwards, increasing its magnitude by about 40\% in the important
asymptotic region outside the nucleus.  This is responsible for the
strong difference between the cross sections calculated with the
vibrational coupling model and the SE method.  The omission
of these coupling effects must cast some doubt on the physical meaning
of the S(2$^+$) and $R_{ce}$ values deduced from the SE method, as
plotted in Fig.~\ref{spectfact_DWBA}.  Experimental spectroscopic
factors S$_{lj}^{exp}$ determined as in Section~\ref{sec:SEcalcs} but
using vibrational formfactors are given in Table~\ref{tab:VIBpar}.
The corresponding $R_{ce}$ ratios, giving the experimental amount of
core excitation, range between 0.17 and 0.24, to be compared with the
minimum value of 0.30 from the standard SE analysis.  The S(0$^+$) and
S($2^+$) values obtained using SE form factors
(Section~\ref{sec:SEcalcs}) are also given (within parentheses) in
Table~\ref{tab:VIBpar} for comparison.
     
\begin{figure}
\centerline{\psfig{file=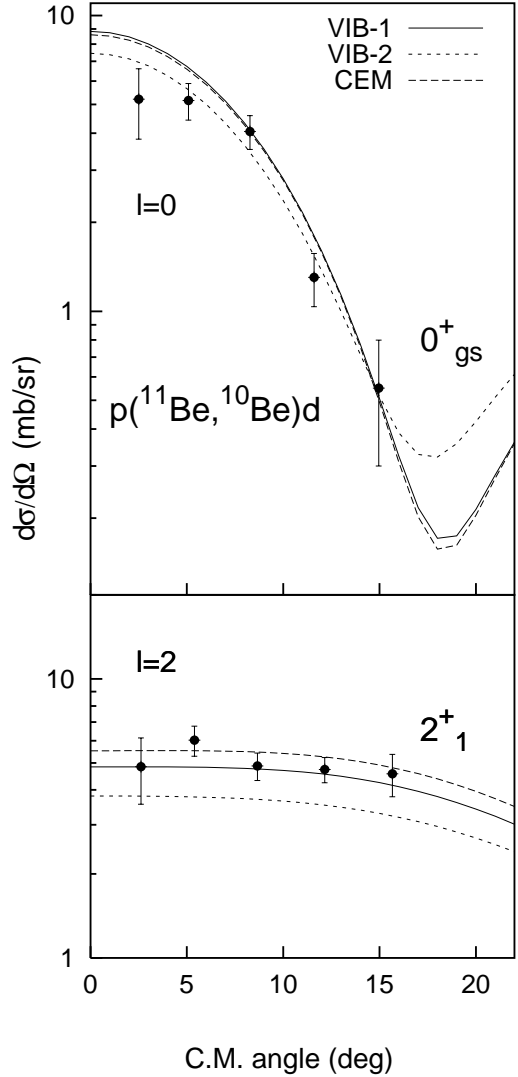,width=0.9\textwidth}}
\caption{Theoretical cross sections for single step transfer 
  with vibrational coupling form factors calculated with different
  geometrical and spin-orbit parameters.  Results obtained with form
  factors from the Core Excitation Model (CEM) of Ref. {\protect
    \cite{Nun96}} are also shown.  The calculations are not normalised
  to the data.}
\label{angdist_vibcoup}
\end{figure}

Alternatively, one can use the unnormalized form factors as
ingredients in the DWBA calculations and directly compare the
predicted cross sections to experimental data. This is done in
Fig.~\ref{angdist_vibcoup}, which shows the results obtained with
optical parameter set $P_3D_3$, and vibrational form factors
corresponding to different well parameters, without any further
renormalisation to the data (i.e.\ the S$_{lj}^{th}$ factors
are implicitly used in these calculations).  As was the case
for the SE method in Section~\ref{sec:SEcalcs}, the calculated cross
sections agree well with the revised last angle data point in the 2$^+$
distribution.  Note that the calculated 2$^+$ cross sections are not
proportional to the theoretical spectroscopic factors, since the VIB-1
and VIB-2 cross sections differ by a factor 1.5 for the same
S$_{d_{5/2}}^{th}$ value (0.16).  The best agreement between
experimental and theoretical cross sections for the 2$^+$ state is
obtained using form factors VIB-1, calculated with the large
spin-orbit parameter ($V_{so}$ = 15.4 MeV) needed to reproduce the
energy of the excited 1/2$^-$ state with the same Woods-Saxon well.
By contrast, the VIB-1b and VIB-2b results (with standard spin-orbit
strengths and $\beta_2$ values of $\sim 0.9$) underestimate the $2^+$
cross section by a factors of 1.4 and 1.9, respectively.

Direct transfer calculations were also performed with radial form
factors from the core excitation model (CEM) of Nunes et al.~\cite{Nun96}. 
This model assumes a rotational structure for the $^{10}$Be
core and possible population of the 2s$_{1/2}$, 1d$_{5/2}$ and
1d$_{3/2}$ by the valence neutron.  We have taken spectroscopic
factors of 0.85, 0.13 and 0.02 for these levels, respectively
(parameter set Be12-b from Ref.~\cite{Nun96a}).  The predicted cross
sections, again with the optical parameter set $P_3D_3$, are shown as
dashed lines in Fig.~\ref{angdist_vibcoup}.  They are rather similar
to the results of the vibrational form factors (VIB-1) and reproduce
the data fairly well.

The $R_{ce}$ value deduced from the present analysis of the
p($^{11}$Be,\,$^{10}$Be)d cross sections therefore provides evidence
for a dominant ($\sim 80\%$) coupling of the halo $1/2^+$ ground state
of $^{11}$Be to the $0^+$ ground state of $^{10}$Be.  The absolute $gs
\rightarrow gs$ spectroscopic factor depends on the geometry adopted
for the neutron well, as might be expected, but it was found not to be
substantially modified by the channel coupling treatment of form
factors (unlike the case of the $2^+$ transition).  Considering the
uncertainties from optical potentials alone, the present (p,\,d)
experimental value (0.67 or 0.80, for geometrical parameters 1 or 2,
respectively), may be considered in good agreement with the
theoretical predictions S$_{s_{1/2}}^{th}$ of the vibrational coupling
model in Table~\ref{tab:VIBpar}.
    
Both of our experimental and theoretical results for the ground state
transition are also in good agreement with the results of two previous
(d,\,p) experiments which used radioactive $^{10}$Be targets.  These
experiments gave spectroscopic factors of 0.73 (Ref.~\cite{Aut70}) and
0.77 (Ref.~\cite{Zwi79}) for incident deuteron energies of 12~MeV and
25~MeV, respectively.  One can also compare the (d,\,p) values for the
transition between the $^{10}$Be$_{gs}$ and the first $1/2^-$ state at
0.32~MeV in $^{11}$Be, 0.63 (Ref.~\cite{Aut70}) and 0.96
(Ref.~\cite{Zwi79}), with our calculated values S$_{p_{1/2}}^{th}$
which range between 0.61 and 0.84 (Table~\ref{tab:VIBpar}).

\subsubsection{Contribution of two-step processes
to the 2$^+$ cross section} \label{sec:2step}

\begin{figure}
\centerline{\psfig{file=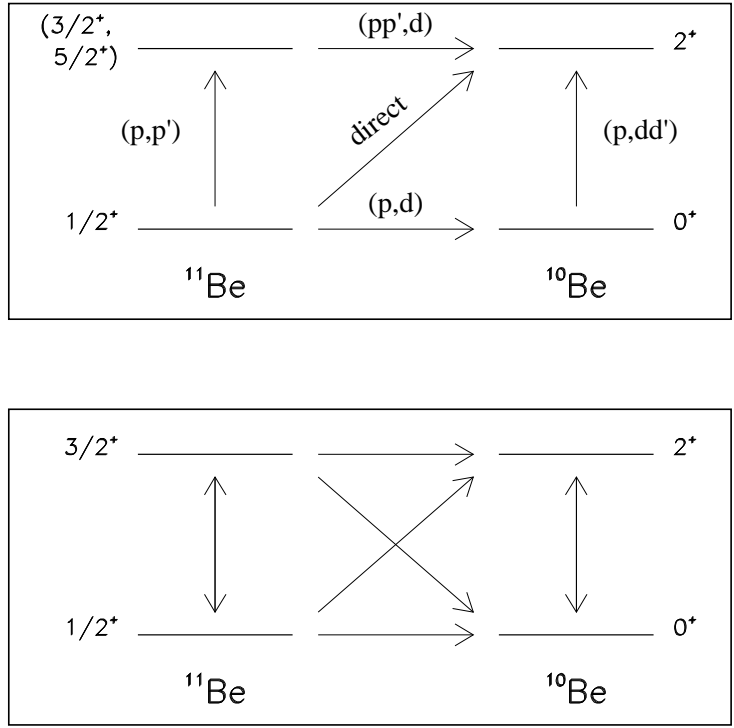,width=0.9\textwidth}}
\caption{Coupling schemes for calculations involving inelastic
  excitation.  Top panel: scheme for the ZR one-way coupling calculations
  leading to the $2^+$ state in $^{10}$Be.  Bottom panel: scheme for
  the full FR-CRC calculations discussed in Section~\ref{sec:CRC} }
\label{coupling_2step}
\end{figure}

An important question is whether processes involving inelastic
excitation in $^{10}$Be and $^{11}$Be could contribute to modify the
2$^+$ cross section in a significant way, as did the coupling in the
form factor.  This has been checked by calculating two-step amplitudes
for inelastic excitation followed by transfer (or vice-versa) with the
zero range code {\sc Cczr}~\cite{vdw90}, which are then added to the 
direct
transfer amplitude (with the appropriate phases).  The coupling scheme
considered in these calculations is shown at the top of in
Fig.~\ref{coupling_2step}.

\begin{figure}
\centerline{\psfig{file=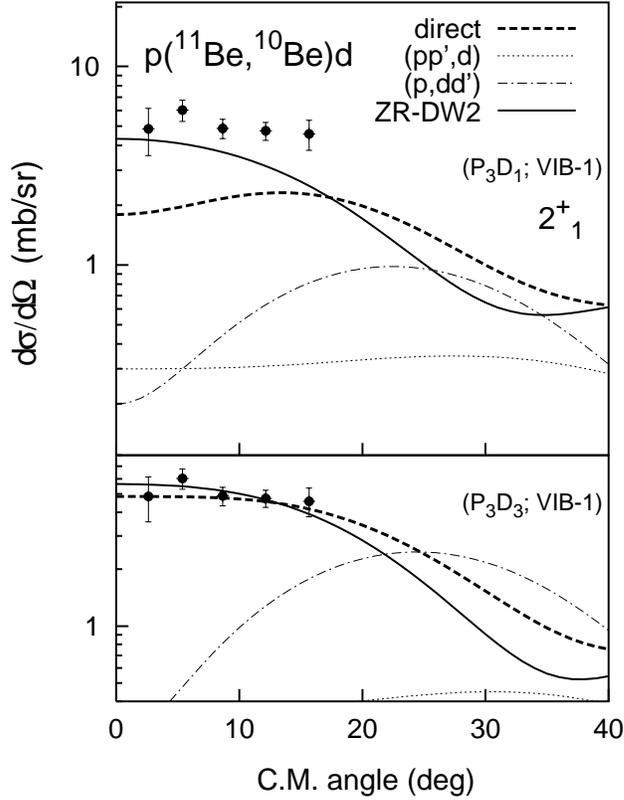,angle=-90,width=0.9\textwidth}}
\caption{Cross sections for the $2^+$ state in $^{10}$Be 
  calculated with {\sc Cczr}.  ``direct'' is for the transfer route
  alone (no coupling), 
  ``(pp$^\prime$,d)'' and ``(p,\,dd$^\prime$)'' are for the
  individual two-step routes (including inelastic one-way coupling),
  ``ZR-DW2'' is for both the two-step routes together with the
  direct.  Results for vibrational form factor VIB-1 and two different
  optical parameter sets are shown.} \label{angdist_2step}
\end{figure}

Only one-way coupling was considered for inelastic excitation, as the
inelastic coupling strength $\beta_2$R was taken equal to the
experimental value in $^{10}$Be \cite{Aut70} extracted with this
assumption.  Vibrational coupling form factors VIB-1 have been used
for the s$_{1/2}$ and d$_{5/2}$ neutron transfers. In addition to the
two-step (p,\,dd$^\prime$) excitation of the 2$^+$ state via the 0$^+$
ground state of $^{10}$Be, possible two-step (pp$^\prime$,d) paths via
3/2$^+$ and 5/2$^+$ states in $^{11}$Be with [2$^+\otimes$s$_{1/2}$]
configuration, were also considered.  These 3/2$^+$ and 5/2$^+$ states
are predicted to have an excitation energy similar to that of the
$^{10}$Be 2$^+$ state (3.37 MeV) by the weak coupling limit of the
vibrational model.  Experimentally, good candidates for these weak
coupling partners of the 2$^+$ state are the unbound levels at 2.69
and 3.41 MeV, with possible J$^{\pi}$ values 3/2$^+$ or
5/2$^+$~\cite{ajz90}.  Inelastic coupling strengths and s$_{1/2}$
neutron form factors of the (p,\,dd$^\prime$) path were also adopted for
(pp$^\prime$,d), with the appropriate phases and amplitudes.  Results
are shown in Fig.~\ref{angdist_2step} for optical parameter sets
$P_3D_3$ and $P_3D_1$, which gave a good description of the 0$^+$
differential cross sections calculated with direct transfer.

For calculations involving inelastic scattering explicitly, where the
macroscopic model of a deformed scattering potential is used for the
inelastic excitation form factor, the standard d~+~$^{10}$Be optical
potential $D_1$ is suitable.  On the other hand, adiabatic deuteron
potentials are not intended for this application, as these potentials
a priori reproduce neither the relevant elastic nor inelastic
scattering channels.  The results labelled $P_3D_3$ in
Fig.~\ref{angdist_2step} have therefore been obtained with the
adiabatic deuteron optical potential $D_3$ as the generator of 
the distorted waves
in the (d+$^{10}$Be) channel, whilst the 
inelastic form factors were proportional
to d$V$/d$r$, the derivative of the standard deuteron
optical potential $D_1$
\cite{Per63}.  

The enhancement of the 2$^+$ cross section at forward angles from the
two-step processes considered here, strongly depends on the optical
parameter sets used in the calculations, as they induce large
differences in both the direct transfer and the inelastic scattering
amplitudes.  For $P_3D_3$, interferences with two-step processes
modify the slope of the angular distribution, but do not change the
cross section at forward angle by more than 15\% relative to single
step transfer, so that it is still in agreement with the present data.
For $P_3D_1$, it is seen in Fig.~\ref{angdist_2step} (top panel) that
interferences with the (pp$^\prime$,\,d) and (p,\,dd$^\prime$) paths
induce a large enhancement of the 2$^+$ cross section (by a factor of
2.4 at 0$^\circ$) relative to direct transfer.

\subsubsection{Comparison of full with one-way coupling calculations
and effect on the 0$^+$ cross section}  \label{sec:CRC}
   
Further coupled-reaction-channels (CRC) calculations have been
performed with the finite-range program {\sc Fresco} \cite{FRESCO}.
These were intended partly as a test of the approximation of
restricting the coupling to one-way, as done in the calculations of
the previous section.  In addition, since the de-excitation of the
$2^+$ level is now considered, the effect of dynamical coupling on the
$0^+$ cross section can be examined.  Finite-range calculations,
however, are not compatible with the Johnson-Soper ADBA approach which
takes into account deuteron breakup effects, since the adiabatic
potential used is obtained from the three-body (core-proton-neutron)
Schr\"odinger equation assuming zero separation between the proton and
neutron.  Hence the calculations presented in this section use the
standard scattering optical potentials $P_3D_1$.  For simplicity, the
(pp$^\prime$d) path through the hypothetical $5/2^+$ state of
$^{11}$Be was not included in these {\sc Fresco} calculations (that
through the $3/2^+$ state {\em was} considered).  Lastly, all the
calculations in this section included the ``remnant term'' in the
interaction potential, normally assumed to cancel in zero-range
distorted-wave programs (the effect of the inclusion of the remnant
term is discussed below).  Thus, for several reasons, the calculated
two-step angular distributions are not expected to be similar to those
presented in the previous section that used the same optical model
potential combination.  The couplings considered are shown
schematically at the bottom of Fig.~\ref{coupling_2step}

\begin{figure} 
\centerline{\psfig{file=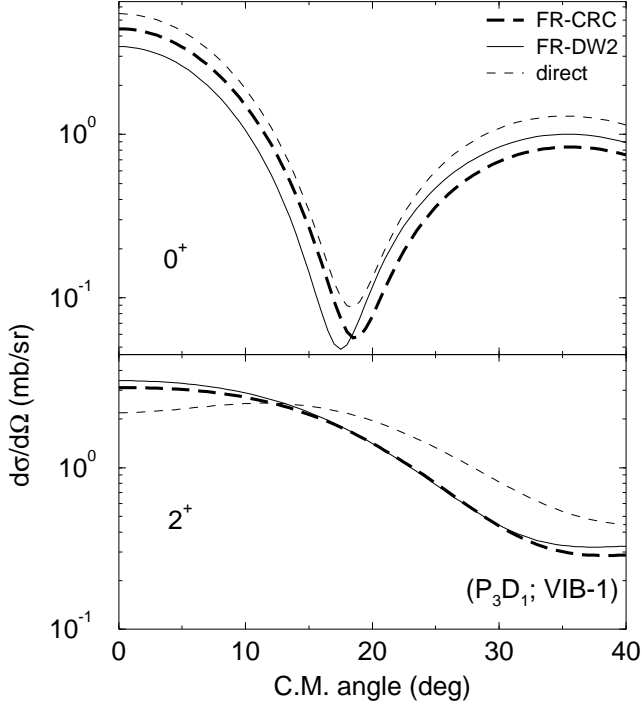,width=0.9\textwidth}}
\caption{Cross sections for the $0^+$ ground
  and $2^+$ 3.37 MeV states in $^{10}$Be calculated with {\sc Fresco}.
  ``FR-CRC'' is for the full finite-range CRC, ``FR-DW2'' is for
  one-way inelastic coupling together with direct transfer, 
 ``direct'' is for the  transfer alone (no coupling).  
  } \label{angdist_10Be_CRC}
\end{figure}

In Fig.~\ref{angdist_10Be_CRC} {\sc Fresco} calculations are presented
for full CRC, for one-way coupling and for the direct transfer alone.
The bottom panel of the figure shows that for the transfer leading to
the $2^+$ state in $^{10}$Be, the difference between the one-way
coupling (FR-DW2) and the full coupling (FR-CRC) cross sections is no
more than 10\%.  As previously seen in the ZR calculations of
Section~\ref{sec:2step}, there is a large difference between the
one-way coupling and the direct transfer cross sections for this
optical model choice.

As regards the transition to $^{10}$Be $0^+$ ground state, the FR
direct, one-way coupling (now including the 2$^+$ to 0$^+$
de-excitation) and full coupled channels angular distributions all
show obvious differences in magnitudes, with the one-way coupling
distribution having in addition a small angular displacement in the
minimum compared to the others (Fig.~\ref{angdist_10Be_CRC}, top
panel).  At small angles, the one-way coupling cross section is about
60\% 
of the direct cross section.  The FR-CRC cross section magnitude is
approximately midway between the two others.  At large angles ($>
20^\circ$), the CRC cross section becomes the lowest of the three.

One should be aware of a possible problem with these comparisons, in
that the same optical potentials have been used for the direct
and coupled-channels calculations rather than refitting the parameters
(and the $\beta_2$ coupling strength) to elastic and inelastic
scattering data with the same model, had such data existed.
Nevertheless, it seems clear that coupling to inelastic channels could
in principle
affect not only the $^{10}$Be $2^+$ and $0^+$ cross sections, but,
more importantly, the ratio $R_{ce}$.

\subsubsection{Effect of the remnant terms in the interaction potential}
   \label{sec:remnant}
   
As mentioned in Section~\ref{sec:CRC} above, one difference between the
finite-range {\sc Fresco} and zero-range {\sc Dwuck} programs, is that
the former allows the inclusion of the full complex remnant terms
(also known as the ``indirect'' part) in the interaction potential.
To be explicit, the prior representation of the interaction for a
pickup reaction $B(b,\,a)A$ with $a = b + x$ is~\cite{Sat83}

\begin{equation} \label{eq:Vint}
  W_\alpha \equiv V_{aA} - U_\alpha =  V_{xA} + (V_{bA} - U_\alpha), 
\end{equation}

where $V_{nN}$ is an effective nucleon-nucleus interaction and
$U_\alpha$ is the distorting potential used for the entrance channel
wave $\chi_\alpha$.  The binding interaction $V_{xA}$ is the only part
from Eq.~\ref{eq:Vint} retained in {\sc Dwuck}, it being argued that
there is considerable cancellation between the ``remnant'' terms
$(V_{bA} - U_\alpha)$.\footnote{In fact the zero-range approximation
  would make the calculation of these terms difficult since each is a
  function of different coordinates.}  However, the cancellation can
never be complete, and the omission of these terms negates the
post-prior equivalence of the interaction.

From a comparison of the results of one-step {\sc Fresco}
calculations, the cross section for the $0^+$ ground state increases
by about 20\% when the remnant terms are included whereas that of the
$2^+$ excited state is not significantly changed.  Thus the {\em
 ratio} of the $2^+$ and $0^+$ cross sections is affected, without
the radial wave functions. The
remnant terms appear to take into account the radial extent of the
$0^+$ wave function being greater than that of the $2^+$.

\subsection{Effect of $^{11}$Be recoil excitation and breakup} 
   \label{sec:REB}

A new method for analyzing (p,d) reactions involving a weakly bound
halo nucleus (C+n) has been developed recently by Timofeyuk and
Johnson~\cite{Tim99} as an extension to the ADBA.  Their adiabatic
approach treats the effect of the recoil excitation and break-up (REB)
of the halo nucleus on the transfer cross section, by replacing the
distorted wave in the p+(C+n) channel by an effective distorted wave,
derived from the potential (p+C) according to a definite prescription.
The distorted wave function in the (d+C) channel is obtained as in the
ADBA, to account for the effect of deuteron break-up on neutron
transfer.

The application of this method to the $^{11}$Be(p,d)$^{10}$Be reaction
at 35 MeV shows that REB enhances forward angle cross sections of the
0$^+$ and 2$^+$ states, leading to an appreciable reduction of
corresponding spectroscopic factors (multiplied by about
0.54 and 0.71, respectively)~\cite{Tim99,Joh99}.  
Such reductions applied to our
results would give a slight increase of $R_{ce}$ and would not change
the lower limit determined for the amount of core excitation in
$^{11}$Be$_{gs}$.  The analysis
method used in Refs.~\cite{Tim99,Joh99} leads to an $R_{ce}$ value close
to 0.5, which falls within the range of values shown in 
fig.~\ref{spectfact_DWBA}.  
However, it has to be observed that the absolute
value of the spectroscopic factor for the 0$^+$ state
obtained by Timofeyuk and Johnson even {\em without} the REB effects,
S(0$^+) = 0.35 \pm 0.04$~\cite{Joh99},
is only in agreement with the lowest of our SE analyses, namely, 
that with the
optical potential $P_1D_3$ (note that refs.~\cite{Tim99,Joh99} used
the SE-1 geometry for the radial form factors).  
Their absolute value for S(0$^+$) of $0.19 \pm 0.02$ {\em with}
REB is difficult to reconcile with, e.g., the measurement of total
reaction cross section~\cite{Tos00} and that for high-energy one-neutron
removal~\cite{Aum00}.

\section{Discussion and Conclusions} \label{secConclude}

The $^1$H($^{11}$Be,$^{10}$Be)$^2$H reaction has been investigated in
order to provide insight on the structure of the ground state of the
halo nucleus $^{11}$Be.  The analysis has attempted to relate the
large cross section observed experimentally for the excitation of the
2$^+$ state at 3.37 MeV with the amount of [2$^+\otimes$1d]
core-excited component in the ground state wave function.

An important ingredient in the calculations is the choice of method
for calculating the radial wave function of the transferred nucleon in
the $^{11}$Be nucleus.  A common choice, which we have labelled SE, is
to calculate the wave function in a Woods-Saxon well, with the depth
adjusted to reproduce the known separation energy.  Another important
ingredient is the use of adiabatic exit channel optical potentials to
approximate the effect of the deuteron breakup.  We find that any
(p,d) reaction analysis using single particle form factors in the SE
method will result in a $^{10}$Be $2^+$ core excitation admixture
$\ge$30\%. 
This result would be in agreement with the predictions of
variational shell model (42\%)~\cite{Ots93} 
although not with most other theoretical models.

However, the validity of the SE single-particle form factors has been
brought into question by coupled-channel calculations in the framework
of the particle-vibration coupling model.  The radial wave function of
the d$_{5/2}$ transferred neutron is found to be strongly modified by
the interaction with the $^{10}$Be deformed core, enhancing the cross
sections relative to the classical SE predictions.  One-step transfer
cross sections calculated with form factors from the present
vibrational coupling approach, or from a core excitation model based
on the assumption of a rotational $^{10}$Be core~\cite{Nun96}, are in
agreement with the data, and predict only $\sim$10 - 20\% admixture of
the [2$^+\otimes$1d] configuration in $^{11}$Be$_{gs}$.  Such a
dominance of the s-wave component is in agreement with many
theoretical calculations (e.g., Refs.~\cite{War92,Nun96,Vin95,Esb95}).

Because of the large deformation of $^{10}$Be, a valid concern is
that coupling channel effects may also contribute to the transfer
cross sections in a significant way, via two-step (inelastic +
transfer) processes.  This has been tested by calculations in
Sections~\ref{sec:2step} and \ref{sec:CRC} using rather simple
assumptions.  The effect of the inelastic coupling could indeed be
significant, but it is dependent on the optical model potentials.  The
breakup of the deuteron seems to have a much more dramatic effect,
even to the extent of changing the conclusions that may be drawn,
regarding the importance of two-step processes. This strongly
indicates that more complete calculations, including a more elaborate
treatment of breakup, 
are required.  These are beyond the scope of
this work.  Effects of $^{11}$Be recoil and break-up should ideally
also be included, as recent calculations~\cite{Tim99} have shown that
they could affect transfer cross sections significantly.  We have also
shown a sensitivity of the calculations to the inclusion of the
remnant term in the interaction potential.  This appears to affect
only the $0^+$ cross section (at the 20\% level), and consequently the
extracted ratio of spectroscopic factors, $R_{ce}$ is also modified.
The necessity of performing both elastic and inelastic scattering
experiments in the (p + $^{11}$Be) and (d + $^{10}$Be) channels at the
appropriate incident energy is evident.  These are needed to fix the
parameters external to the transfer process.  It should also be noted
that it is the nucleon-core potential, p~+~$^{10}$Be, which is required
as input to the ADBA and REB adiabatic calculations.

Three other key experimental results concerning the structure of
$^{11}$Be$_{gs}$ should be considered.  Following a
demonstration by Suzuki et al.~\cite{Suz95} of the sensitivity of the
ground state magnetic moment of $^{11}$Be to the [$2^+ \otimes$
1d$_{5/2}$] component of the wave function, a measurement of the
magnetic moment has been made at ISOLDE and has given a value of
$-1.6816(8) \mu_{\rm N}$~\cite{Gei99}.  This result implies a relatively pure
2s$_{1/2}$ halo state with essentially no 1d$_{5/2}$ admixture, but a
precise interpretation is dependent on the assumed quenching of the
single particle magnetic moment.  Another recent experiment was a
high-energy single-neutron removal (``knockout'')
reaction~\cite{Aum00} which used a  $^{11}$Be beam.  
The authors' measured cross sections to the
particle-bound levels in $^{10}$Be agreed well with
single-particle cross sections calculated in an eikonal
model~\cite{Tos99} multiplied by spectroscopic factors from the shell
model~\cite{War92}.  It was concluded that $^{11}$Be ground state has a
dominant 2s single particle character with a small 1d component. 
The third experiment is the measurement of the total reaction cross
section of $^{11}$Be~\cite{Tan85}, which was one of the earliest 
indicators of the extended halo structure of this nucleus.
Reaction calculations~\cite{AlK96} can interpret this in terms of the
average radius of the halo, which in turn is related to the percentage
of s-wave component in the $^{11}$Be wave function.  The result~\cite{Tos00}
is that the s-wave fraction is of order 0.8 to 0.9, and larger d-wave
admixtures could not be reconciled with the measured cross section.

In conclusion, we have explored the structure of $^{11}$Be ground
state via a single-neutron transfer reaction with a $^{11}$Be
radioactive beam.  From the experimental point of view, the data
presented here constitute a successful implementation of the technique
of transfer reactions in inverse kinematics to study the structure of
light exotic nuclei, as suggested in, e.g.,
Refs.~\cite{Win97,Len98c,Har92,Ege97}.  The present ``best estimate''
of the $^{11}$Be$_{gs}$ wave function is a dominant 2s component with
a 0.16 [$2^+ \otimes $ 1d] core-excitation admixture.  This value may
be better defined by future calculations that principally incorporate
a model of deuteron breakup (as well as that of $^{11}$Be) within a CRC
framework.

\ack

We wish to thank the GANIL technicians for their dedicated assistance
during this experiment.  Natasha Timofeyuk is acknowledged for
communicating the results of EFR-DWBA calculations and Filomena Nunes
for providing $^{11}$Be wave functions from rotational coupling
calculations. We are indebted to Ian Thompson for assistance with the
CRC calculations, and to Jeff Tostevin and Nicole Vinh-Mau for
fruitful discussions.  Financial support from the Centre National de
la Recherche Scientifique (France) and the Engineering and Physical
Sciences Research Council (UK) is gratefully acknowledged.  One of us
(JSW) was supported by a contract from the IN2P3-CNRS at IPN-Orsay and
subsequently an EPSRC Visiting Fellowship (Grant No.\ DMR162/384) at
the University of Surrey during part of the work presented here.

\appendix

\section*{Appendix A: Transfer form factors in a vibrational coupling 
approach}

It has long been established that in the presence of core-polarisation
admixtures created by an interaction Hamiltonian $H_{int}$, radial
form factors of one-nucleon transfer reactions should no more be
approximated by the product of a spectroscopic amplitude and a
single-particle wave function $U_{lj}^{SE}$ deduced from the standard
Separation Energy procedure, but have to be determined by solving the
Pinkston-Satchler coupled equations~\cite{Pin65}. In the present
analysis of the $^{11}$Be(p,\,d)$^{10}$Be reaction, these coupled
equations were solved in the framework of the particle-vibration
coupling model~\cite{BM,vdw96}, with the program CCVIB~\cite{vdw98}.
The interaction Hamiltonian was chosen as:
\begin{equation}
H_{int}=-R {{dV} \over {dr}} \sum_{LM} {{i^{-L} \beta _L} \over{\sqrt{2L+1}} }
\left( c_{LM}^{\dag}+ (-1)^{L+M} c_{L-M} \right) Y_{LM}^{\ast}(\theta,\phi)
\end{equation}
where $c_{LM}^{\dag}$ and $c_{L-M}$ are operators for phonon creation
and annihilation and $\beta_L$ is the deformation parameter.  Here the
term d$V$/d$r$ corresponds to the deformation of the central part
$V(r)$ of the potential, with no action on the spin-orbit part.
Coupled-channel calculations were performed for both 1/2$^+$ ground
state and 1/2$^{-}$ first excited state at 0.32 MeV, with the
additional constraint that the same well depth $V_0$ reproduced the
experimental separation energies relative to the ground state of
$^{10}$Be, thus ensuring the right asymptotic behaviour of the wave
functions.  This reproduction of the separation energies could be
obtained by adjusting either the value of the $\beta_2$ parameter or
the strength of the spin-orbit part of the potential.  The
configuration space was truncated to the 0$^+_{gs}$ and 2$^+_{1}$ core
states in $^{10}$Be (first order vibrational coupling), coupled with
one neutron in the 1p$_{3/2}$ or 1p$_{1/2}$, and 2s$_{1/2}$ or
1d$_{5/2}$ orbitals, for the 1/2$^-$ and 1/2$^+$ states, respectively.

The resulting wave functions 
were normalised for each final state according to the prescription
$\int (\sum_{lj} u_{lj}^2(r))r^2dr=1$.  The relative weights of the
different configurations are given by the corresponding squared
amplitudes $\alpha_{lj}^2=(\int u_{lj}^2(r) r^2 dr)^2$.  Strictly
speaking, spectroscopic factors $S_{lj}$ are defined~\cite{Aus70} as
the product of the squared overlap integral $(\int
U_{lj}^{sp}u_{lj}(r) r^2 dr)^2$ and $(n+1)$, where $n$ is the occupancy
of the orbital $lj$ in the core nucleus and $U_{lj}^{sp}$ is the
corresponding single-particle wave function calculated in the same
Woods-Saxon potential well.  The factors $S_{lj}^{th}$ given
in Table~\ref{tab:VIBpar},
which are in fact the squares of parentage amplitudes although
we call them theoretical spectroscopic factors for convenience,
were calculated by assuming $n = 0$ and, for
the particular cases of s$_{1/2}$ and d$_{5/2}$, by approximating the
unbound single-particle wave functions $U_{lj}^{sp}$, by the bound
$U_{lj}^{SE}$ radial form factors determined by the SE method.  They
were found to differ from the relative weights $\alpha_{lj}^2$ by less
than 1\%.

The radial wave functions $u_{lj}$(r) obtained by solving coupled equations
can be directly  used in the calculation of transfer cross sections
by DWBA and/or CRC methods, so that the comparison of theoretical and
experimental results provides a test of the
validity of the present vibrational coupling model and
other ingredients such as optical potential parameters. Another possible
procedure is to normalise the calculated wave functions
by dividing them by $\int u_{lj}^2(r)r^2 dr$,
and use the renormalised U$_{lj}$(r)  as ``realistic''
form factors in DWBA analysis, thus giving  experimental spectroscopic factors
expected to be more reliable than those extracted by the conventional SE
method. Both methods were used in the present
analysis of the $^{11}$Be(p,\,d)$^{10}$Be reaction.



\end{document}